\documentclass[%
reprint,
superscriptaddress,
amsmath,amssymb,
aps,
prb,
floatfix,
]{revtex4-1}
\usepackage{graphicx}%
\usepackage{amsmath}
\usepackage{amssymb}
\usepackage{subcaption}
\usepackage{graphicx}
\usepackage{dcolumn}%
\usepackage{comment}
\usepackage{bm}%
\usepackage[names]{xcolor}

\begin{document}
\preprint{APS/123-QED}

\title{Finite-temperature materials modeling \\from the quantum nuclei to the hot electrons regime}%

\author{Nataliya Lopanitsyna}
\affiliation{%
 Laboratory of Computational Science and Modeling, IMX, \'Ecole Polytechnique F\'ed\'erale de Lausanne, 1015 Lausanne, Switzerland
}%

\author{Chiheb {Ben Mahmoud}}
\affiliation{%
 Laboratory of Computational Science and Modeling, IMX, \'Ecole Polytechnique F\'ed\'erale de Lausanne, 1015 Lausanne, Switzerland
}%

\author{Michele Ceriotti}
\email{michele.ceriotti@epfl.ch}
\affiliation{%
 Laboratory of Computational Science and Modeling, IMX, \'Ecole Polytechnique F\'ed\'erale de Lausanne, 1015 Lausanne, Switzerland
}%

\date{\today}%

\begin{abstract}

Atomistic simulations provide insights into structure-property relations on an atomic size and length scale, that are complementary to the macroscopic observables that can be obtained from experiments. Quantitative predictions, however, are usually hindered by the need to strike a balance between the accuracy of the calculation of the interatomic potential and the modelling of realistic thermodynamic conditions. 
Machine-learning techniques make it possible to efficiently approximate the outcome of accurate electronic-structure calculations, that can therefore be combined with extensive thermodynamic sampling.
We take elemental nickel as a prototypical material, whose alloys have applications from cryogenic temperatures up to close to their melting point, and use it to demonstrate how a combination of machine-learning models of electronic properties and statistical sampling methods makes it possible to compute accurate finite-temperature properties at an affordable cost. We demonstrate the calculation of a broad array of bulk, interfacial and defect properties over a temperature range from 100 to 2500 K, modeling also, when needed, the impact of nuclear quantum fluctuations and electronic entropy. 
The framework we demonstrate here can be easily generalized to more complex alloys and different classes of materials. 

\end{abstract}

\maketitle

\section{\label{sec:level1}Introduction}

Computational modelling has been used for several decades to gain qualitative, mechanistic understanding into the atomic-scale phenomena that underlie the structure-property relations in materials.\cite{yip05book} The rise of comparatively affordable electronic-structure calculations based on density functional theory~\cite{parr-yang94book} has made it possible to achieve predictive accuracy for several structural, mechanical and functional properties, assisting the design and optimization of materials for both fundamental and technical applications~\cite{ceder1998identification, besenbacher1998design, greeley2006computational,yan2015material, chang1984structural, oganov2009quantify,pickard2011ab}. 

However, the vast majority of existing calculations have been performed in terms of static structures, corresponding to minima in the potential energy of the system, possibly with local free-energy corrections based on a harmonic approximation~\cite{olsson2014ab,zhang2009thermodynamic, koba+17prm, minakov2015melting, zhang2009thermodynamic}. To achieve predictive accuracy across the entirety of the phase diagram of a material, one has to incorporate accurate finite-temperature thermodynamics, which involves dealing with the quantum mechanical nature of the nuclei when below the Debye temperature, with a full treatment of anharmonic fluctuations when approaching the melting point, and with the modelling of electronic excitations beyond the Born-Oppenheimer approximation at even higher temperatures. 
Immense efforts, led among others by J\"org Neugebauer and collaborators~\cite{glensk2015understanding,ishibashi2020correlation,glensk2014breakdown} have shown how it is possible to obtain accurate thermodynamics of materials -- metals in particular -- based on first-principles electronic structure calculations. These efforts have also shown how a description of electronic (and magnetic) excitations is also necessary to fully account for the thermophysical properties of materials, e.g. their heat capacity. 

The advent of data-driven approaches to build accurate, yet affordable, interatomic potentials based on the regression of energy and forces from reference electronic structure calculations have reduced considerably the effort needed in evaluating the finite-temperature thermodynamics of materials with first-principles accuracy, which has made it possible, for instance, to investigate the finite-temperature mechanical properties of iron~\cite{drag+18prm} to determine the subtle difference in free energy between different phases of water~\cite{chen+19pnas}, or to study the phase diagram of hybrid perovskite materials~\cite{jinnouchi2019phase}. 
Here we demonstrate the combination of machine-learning potentials with thermodynamic integration and finite-temperature sampling to compute bulk and interfacial properties of materials from cryogenic temperatures up to above the melting point. 
We also use a recently-developed scheme to predict the electronic density of states to take into account the impact of electronic excitations, without the need to perform additional electronic-structure calculations. 
We use nickel as a test system, a metal whose alloys find applications across a very broad temperature range, and which is sufficiently well studied to provide reliable reference data for most of the properties we consider. 
In Section~\ref{sec:methods} we summarize the details of the reference calculations, and the construction of the ML models; in Section~\ref{sec:results} we demonstrate the accuracy of the machine-learning potential, and we compute several challenging finite-temperature properties of Ni, comparing the results of the ML model, of a state-of-the-art empirical forcefield~\cite{pun_development_2009}, and of experiments and DFT when available. Finally, we draw our conclusions.

\section{\label{sec:methods}Methods}
We begin by  describing the details of the underlying reference electronic-structure calculations, the construction of the training set, and the structure of the machine-learning model.

\begin{figure}[bthp]
    \includegraphics[width=1.\columnwidth]{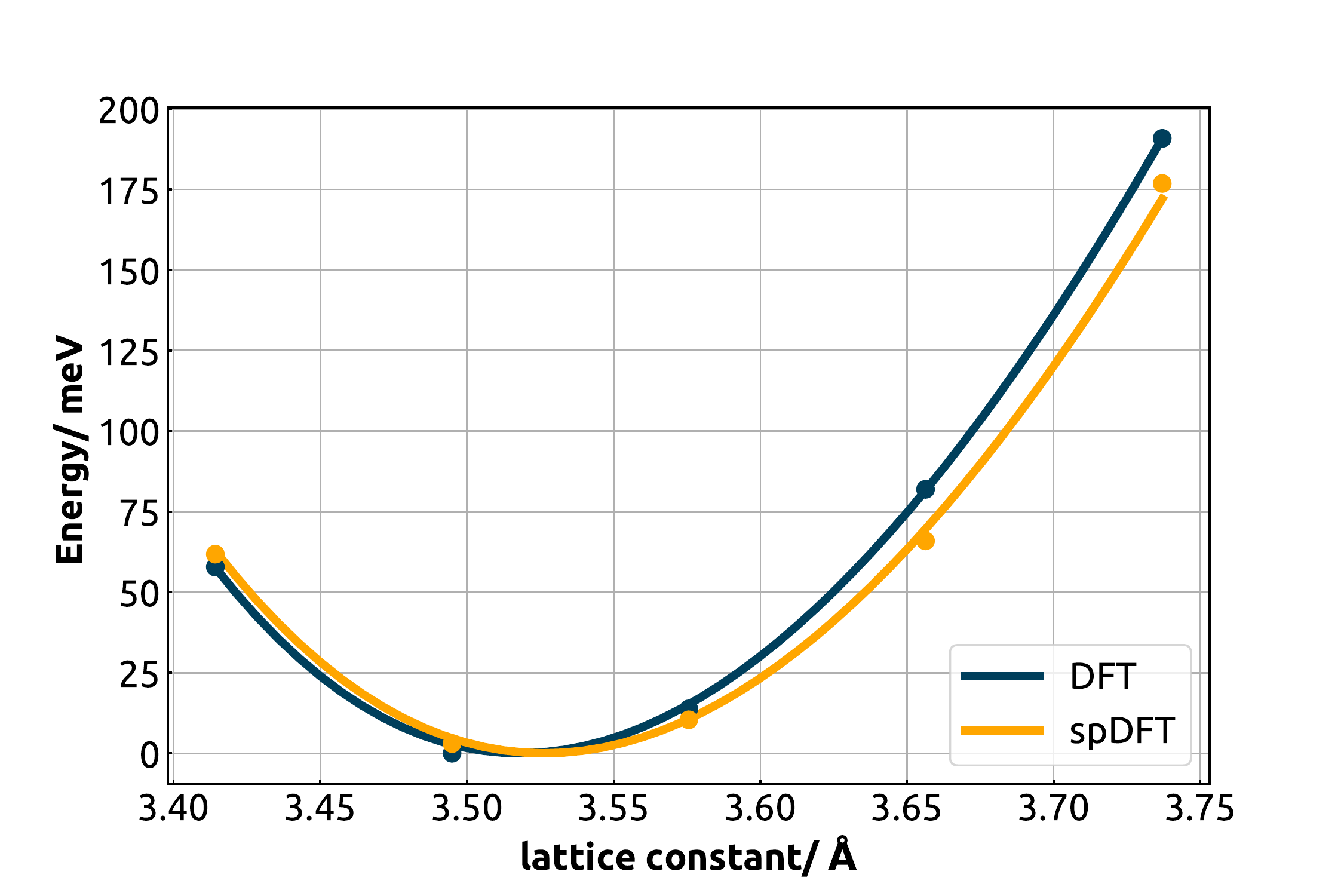}
    \caption{Equation of state of FCC nickel, (referenced to the minimum energy): blue dots represent DFT calculations, yellow dots spin-polarized DFT (spDFT) calculations. Solid lines indicate the corresponding fits to a Birch-Murnaghan equation of state.
    }
    \label{fig:evsv}
\end{figure}

\subsection{Electronic structure reference}

We compute all the energies and forces using   density-functional theory (DFT), as implemented in QUANTUM ESPRESSO \cite{giannozzi2009quantum}. We use the PBE exchange-correlation functional \cite{perdew1996generalized}, together with an ultrasoft pseudopotential \cite{laasonen1991implementation} with 10 valence electrons for Ni, from the standard solid-state pseudopotential library~\cite{pran+18npjcm}. The wave function is expanded in plane waves with a cutoff energy of 40Ry. The Brillouin zone sampling uses the Monkhorst-Pack scheme \cite{monkhorst1976special} with a k-point density of 0.07 \AA$^{-1}$. To improve the convergence of the integral over the k-points mesh, we use the Methfessel-Paxton first-order spreading~\cite{methfessel_high-precision_1989}  with a broadening parameter equal to 0.0441 Ry. All the parameters are kept fixed for the whole data set and are converged in terms of energy differences.   
We perform all the reference calculations with non spin-polarised DFT, even though Nickel exhibits ferromagnetic ordering below its Curie temperature (628K). This choice is driven by the fact that treatment of magnetism implies associating additional degrees of freedom describing the magnetic configuration of the system (e.g. spin polarization of atoms for a collinear treatment), that is not compatible with the typical infrastructure of machine-learning potentials, that use only nuclear coordinates as inputs. 
As an approximate alternative, one would need to perform a separate set of reference calculations below and above the Curie temperature, which however would introduce an undesirable temperature dependence of the potential.
Given that our main goal is to describe high-temperature conditions, where anharmonic contributions to the free energy become important, we prioritize the description of the paramagnetic phase. Furthermore, it
has been shown that the magnetism has a limited impact on the properties of Ni, and it doesn't introduce a significant difference for phonon dispersion curves, vacancy formation energies, thermal expansion \cite{Wang_2004,K_rmann_2016, gong_temperature_2018}. Indeed, as shown in Figure \ref{fig:evsv}, the equation of state computed with non spin-polarized DFT (blue curve) and collinear spin-polarized DFT (yellow curve, ferromagnetic ordering) exhibits very small differences. The lattice constant changes by less than 1$\%$   (3.517\AA{} for DFT vs 3.526 \AA{} for spDFT) and bulk moduli differ by 5$\%$ (195$\mathrm{GPa}$ for DFT and 185$\mathrm{GPa}$ for spDFT).  

\begin{table}[btp]
\begin{tabular}{lll}
\hline
\hline
 Structure type & No. structures & No. atoms  \\ \hline
  Selected from REMD & 988  & 108   \\
  FCC Bulk & &\\
  \hspace{0.5cm} isotropic stress & 4  & 108   \\
  \hspace{0.5cm} shear stress & 4  & 108   \\
  \hspace{0.5cm} uniaxal stress & 4  & 108   \\
  \hspace{0.5cm} displacement of one atom &10 & 108\\
  Single vacancy& 6& 107 \\
  Single intestitial & 18& 109\\
  HCP Bulk & 22  & 54 \\
  BCC Bulk& 10 & 54 \\
  Stacking fault &299 &24\\
  Solid-Vacuum Interface & &\\
\hspace{0.5cm}(100)AA & 154 & 9 \\
\hspace{0.5cm}(100)AB & 110 & 8 \\
\hspace{0.5cm}(110)AA & 88  & 13\\
\hspace{0.5cm}(110)AB & 252  & 12\\
\hspace{0.5cm}(111)AA & 110 & 8 \\
\hspace{0.5cm}(111)AB & 105 & 9 \\
Solid-Liquid Interface & 17 & 96 \\
Liquid-Vacuum Interface &10 & 108 \\
Other &24 & 7 \\
  Total & 2235 & -- \\\hline
\end{tabular}
\caption{Overview of the composition of the training dataset used to fit the neural network potential. The first column shows the number of structures included in each group, and the second column shows the number of atoms included in each supercell. }
\label{tabl:dataset}
\end{table}

\subsection{Training set construction}

The structures included into the dataset are selected by an iterative procedure, building on the insights of existing literature in similar systems \cite{drag+18prm,koba+17prm,bartok2018machine}. Each training structure provides total energy and forces computed by DFT, with the computational settings indicated above. In Table~ \ref{tabl:dataset} we summarize the content of the final version of the dataset. 
Given the availability of a reliable EAM potential\cite{pun_development_2009}, we  obtain a set of ~1000 structures selected by farthest point sampling (FPS)~ \cite{ceriotti2013demonstrating} out of a replica exchange molecular dynamics simulation \cite{Sugita1999}, performed with the i-PI implementation \cite{chen_improving_2015,kapi+19cpc}, and including 82 $NpT$ trajectories spanning a broad range of temperatures (100K - 3200K) and pressures(-5GPa - 5GPa).
 
On top of these baseline structures, that provide a diverse set of configurations across the phase diagram of Ni, we incorporate targeted single-point calculations that are used to ensure that configurations that are relevant for important structural and mechanical properties are well represented. %
In particular,  we include 1x1x1 FCC structures stretched and compressed by less than 3\% of the equilibrium lattice parameter -- that report directly on bulk modulus and elastic constants. To reproduce accurately defects formation energies, we perform geometry optimization of a single vacancy and interstitial in 3x3x3 FCC cells at 0K, and of a 3x3x1 HCP and 3x3x3 BCC cells, describing metastable phases of Ni. We also include 1x1x6 FCC structures with the x-, y- and z-axes oriented along $[11\bar{0}], [11\bar{2}]$, and [111] directions, distorted to incorporate information on the generalized stacking fault surface, as well as 
(111), (001), and (110) surfaces created by rigid cleavage of the bulk, leaving a slab which is more than 12\AA{} thick. All the steps of the geometry optimization have been added to the reference set.
Finally, to ensure that the NNP samples the repulsive part of the interatomic potential, we add 3x3x3 FCC structures where a position of one atom has been randomized by up to 0.5 - 1.2 $\AA$.
In total, the training set contains approximately 2200 configurations.

\begin{figure}[tbp]
    \includegraphics[width=1.05\columnwidth]{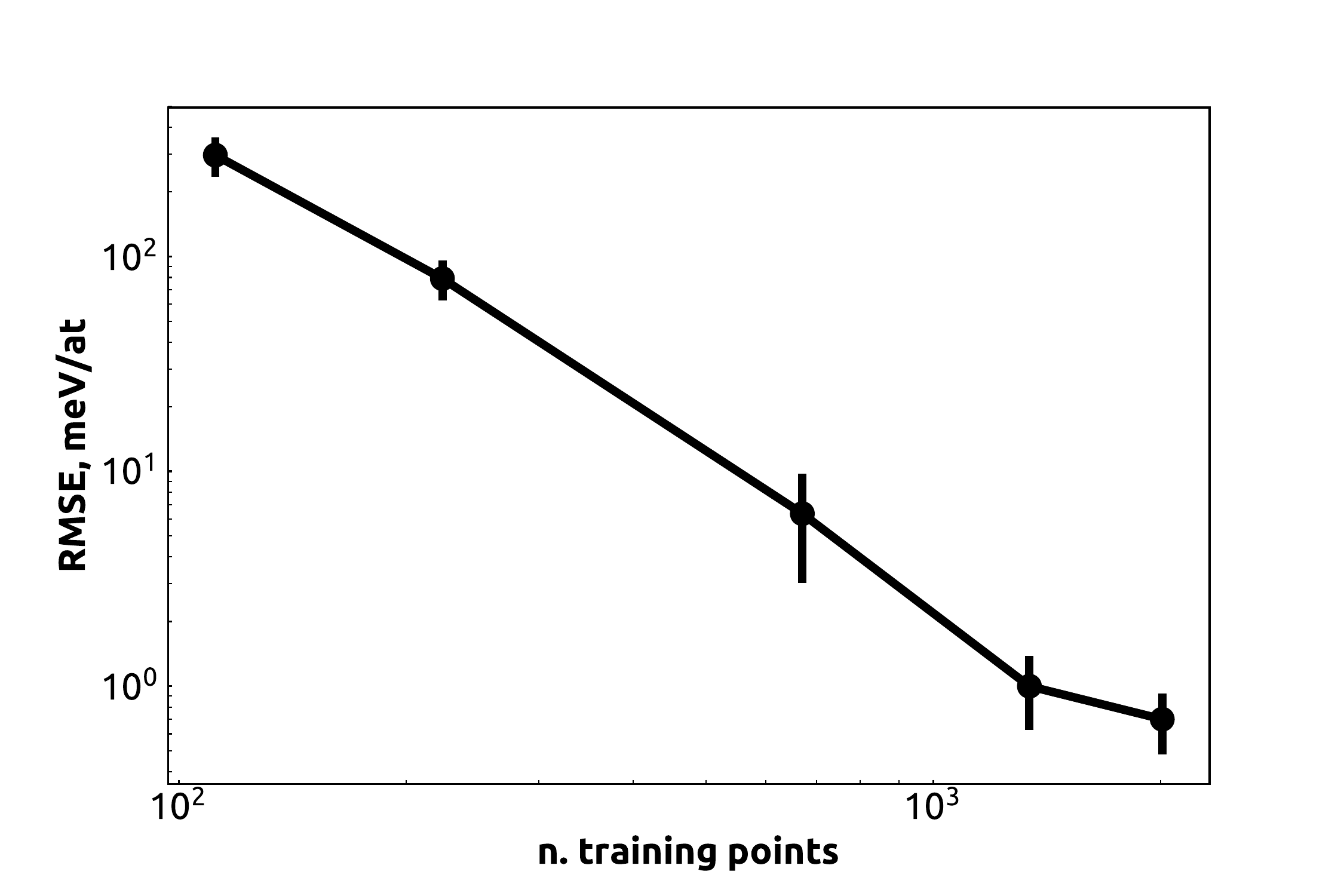}
    \caption{Learning curve for the energy RMSE of the NN potential, as a function of the number of structures included in the training set. The points and error bars indicate the mean and standard deviation of five potentials, computed with different random choices of the training points. }
    \label{fig:lc}
\end{figure}

\subsection{Neural network potential}

To describe the interatomic potential we train a high-dimensional neural network (NN) following the approach proposed by Behler and Parinello \cite{behl-parr07prl}. Since the framework of the Behler-Parinello neural network potentials has been discussed extensively in the literature \cite{behl-parr07prl, artr-behl12prb, artr+11prb,behl11pccp}, we only outline briefly the fundamental underlying concepts. 
 
\newcommand{\natoms}{N_\text{atoms}}
The total energy $E$ of a structure $A$ with $N$ atoms could be represented as a sum of atomic energy contributions associated with environments $A_i$ centered on each atom:
\begin{equation}
    E(A) = \sum^{\natoms}_n E(A_i)
    \label{eq:etot}
\end{equation}
The expression of the atomic energy contributions is written as a series of nested activation functions acting on linear combinations of the values in the previous layer. The input layer, that describes the geometry of each atom-centred environment, entails a vector of atom-centred symmetry functions, that describe two and three-body correlations between neigbours~\cite{behl11jcp,will+19jcp}.
The architecture of the NN and the functional form of the symmetry function are analogous to those used in Ref.~\citenum{imba+18jcp}. The values of the parameters defining the set of symmetry functions were determined by first generating a large set of possible features, and selecting the most informative ones based on the CUR algorithm, as discussed in Ref.~\citenum{imba+18jcp}.
The parameters of the network are optimized using the N2P2 package\cite{singraber2019parallel,N2P2} to agree with the reference DFT data. The resulting parameters of the potential are given in the SI.
The NN architecture includes 2 hidden layers with 25 nodes each. 50 symmetry functions are selected with CUR~\cite{maho-drin09pnas} out of an initial pool combining cutoff distances of 8, 12, 16 and 20 Bohr. 90\%{} of the dataset set is used for training, with a random selection including 10\% of structures being held out for validation. The RMSEs on the training and testing subsets are 0.45{meV/atom} and 0.55{meV/atom} for energies and 22{meV/\AA} and 23{meV/\AA} for forces respectively.  These errors -- as well as the errors on selected target properties, discussed in Section~\ref{sec:results} -- are in line with state-of-the-art potentials, and comparable with the typical error of density functional theory. As shown in Figure~\ref{fig:lc}, the model accuracy is limited by the amount of training data, and not by the complexity of the model, so it would be easy, if needed, to further reduce the error by just increasing the train set size.

\begin{figure}
    \includegraphics[width=1.\columnwidth]{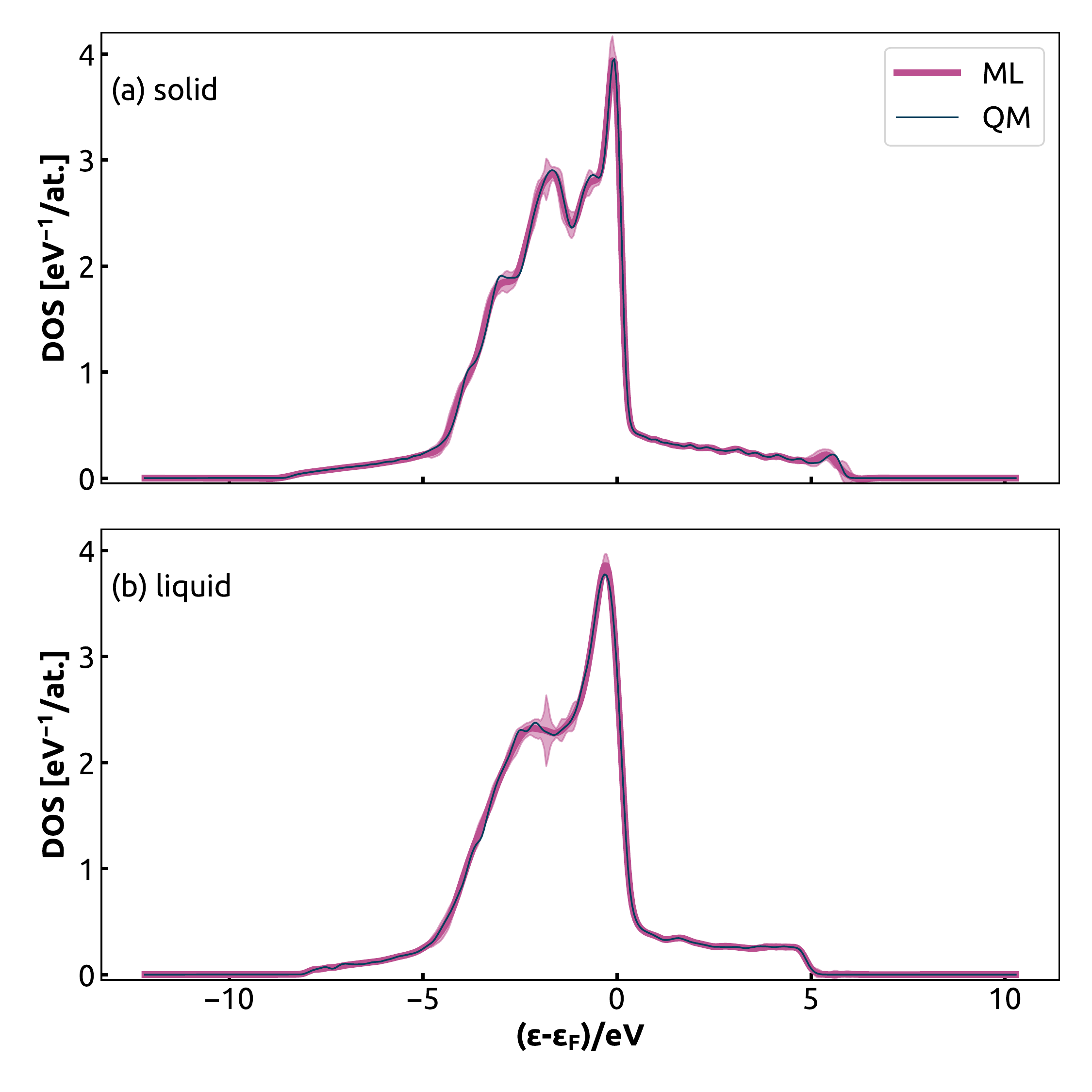}
    \caption{Two examples of the DOS prediction, from the pointwise representation, from the Nickel data set, where (a) is a bulk FCC structure and (b) is a liquid structure. Both predictions are compared to their respective DFT reference. We use a committee model of 8 GPR models. The shaded area represents the uncertainty at every energy level. The zero level is the Fermi energy of the reference structures. The target DOS is generated with a Gaussian broadening of $0.1$eV} %
    \label{fig:dos_examples}
\end{figure}

\begin{figure}
    \includegraphics[width=1.\columnwidth]{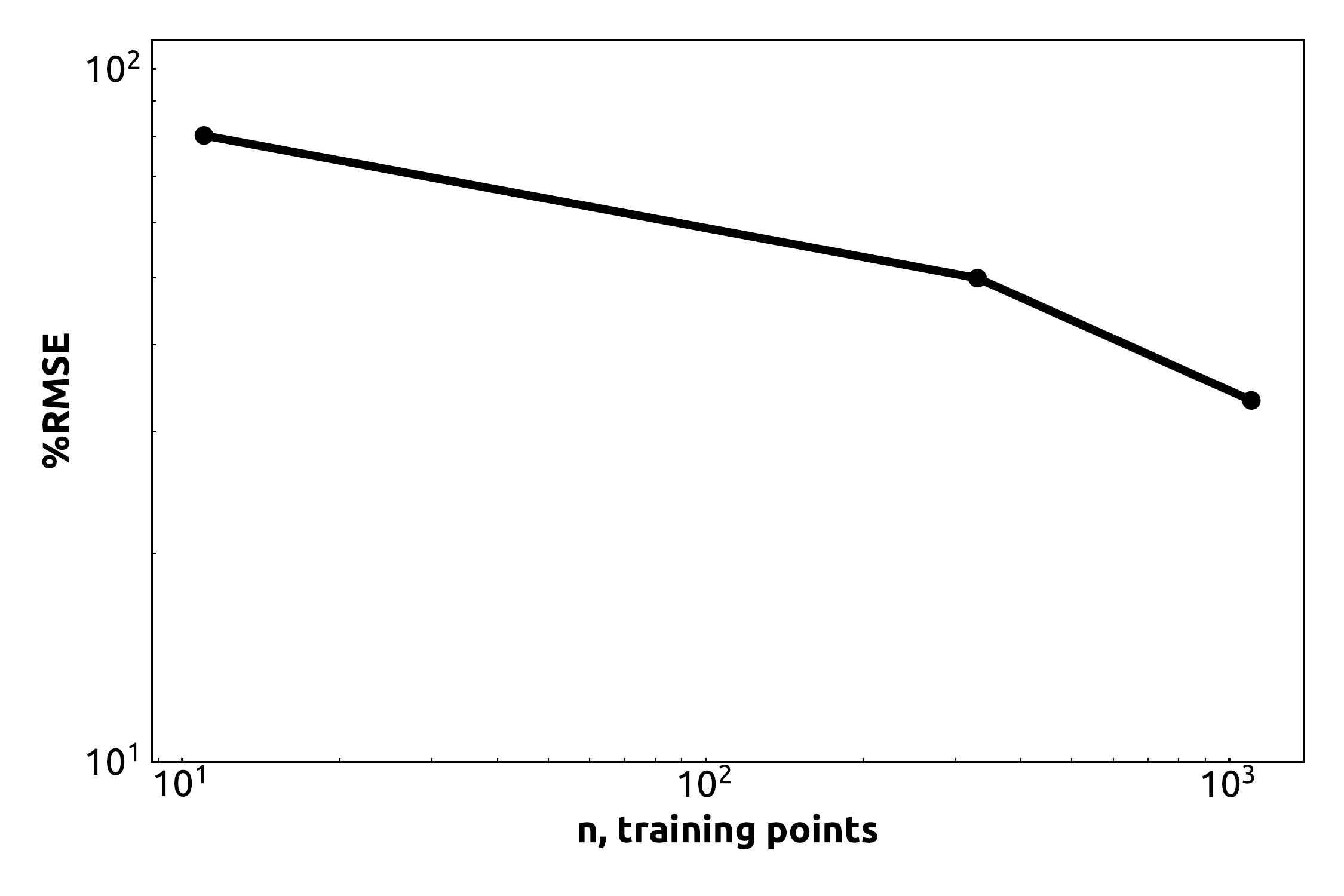}
    \caption{Evolution of the prediction errors as a function of the training set size, using the pointwise representation of the ML DOS. The reference DOS is generated with a Gaussian broadening of $0.1$eV}
    \label{fig:lc_dos}
\end{figure}

\subsection{\label{sec:dos-theory} ML model of the electronic density of states}

A NN potential allows to sample phase space in a way that is consistent with ab initio quality energetics. However, it does not give direct access to electronic-structure properties. 
Recently, ML models have been proposed that give direct access to properties that are related to the electronic degrees of freedom, such as the ground-state charge density~\cite{broc+17nc,gris+19acscs,fabr+19cs} and the density of single-particle energy levels (density of states, DOS)~\cite{benmahmoud2020arxiv}. 
As a first step towards a fully integrated, universal ML scheme that provides a complete surrogate model of quantum mechanical calculations, we train a DOS model and we use it to predict properties that depend on electronic excitations, such as the high-temperature heat capacity.

We use an atom-centered model for the DOS, where we expand the total DOS of a structure A over a sum of local DOS contributions (LDOS) associated with its atomic environments A$_{i}$:
\begin{equation*}
    \mathrm{DOS}(A, E) = \sum_{i \in A} \mathrm{LDOS}(A_i, E).
\end{equation*}
The reference DFT DOS is constructed with a Gaussian broadening $g_{b}=0.1$eV, which ensures that the curves are well-detailed. We use the Fermi energy $\varepsilon_F$ of each structure as the energy reference.

We follow the approach introduced in Ref.\citenum{benmahmoud2020arxiv} to determine the mapping between the atomic environment $A_i$ and its contribution to the total DOS. In a nutshell, we introduce a positive-definite scalar kernel $k(A_i, A'_{i'})$ that describes the similarity between two atomic environments.
We use in practice the SOAP kernel~\cite{bart+13prb}, as implemented in librascal~\cite{LIBRASCAL}. 
We then determine the \textit{active set} containing the $M$ most diverse environments found in the training set, and write a Projected Process (PP) approximation of the Gaussian Process (GP) algorithm  to express the LDOS as a function of the basis set formed by the kernel between each target environment and the active set
\begin{equation*}
    \mathrm{LDOS}(A_i, E) = \sum_{j \in M} x_j(E) k(A_i, M_j).
\end{equation*}
Note that the expansion coefficients  $\mathbf{x}_M(E)$ are determined separately for each energy channel.
We use the pointwise representation of the DOS from Ref.\citenum{benmahmoud2020arxiv}, where we discretize the energy axis over a finite range and take the DOS at every energy point as a target of the ML model. 
Once the model is trained, the DOS of a new structure $A_*$ can be easily obtained from the dot product between the kernel matrix of its atomic environments and the active set, and the energy-dependent expansion coefficients $\mathbf{x}_M$. 
To monitor the reliability of the predictions, we also implement uncertainty estimation based on a calibrated committee model~\cite{musi+19jctc}.

To train a model of the DOS we use a subset containing 1377 structures of the data set in Tab.~\ref{tabl:dataset}, discarding those corresponding to Solid-Liquid, Liquid-Vacuum and Solid-Vacuum interfaces. We use the radial cutoff $r_0=6.$\AA{}  and an atomic density smoothing $\sigma_{at}=0.45$ for the SOAP features. 
The active set contains $15000$ environments selected by FPS out of the $\approx 130000$ that are present in the training set. We determine the regression weights $\mathbf{x}_M$ using a regularization parameter that is optimized by a 10-fold cross-validation scheme, in order to ensure the the model is not in the over-fitting regime.

The normalized prediction error from the cross-validated model, computed by dividing the integrated root mean square error (RMSE) of the ML DOS and DFT DOS by the integrated standard deviation of the reference DFT DOS, is $24\%$. While this error might seem large, in practice this level of error is sufficient to estimate key properties for the electronic contributions, such as the density of states at the Fermi level. 
Fig.~\ref{fig:dos_examples} shows two representative examples of the predictions from DOS ML model for the DOS of solid and liquid Nickel, compared with the reference DFT DOS. It is clear that the ML DOS reproduces qualitatively and quantitatively the shape of the DOS in the two phases. As shown  Fig.~\ref{fig:lc_dos}, the learning curve is far from saturation, and a more accurate model could be obtained, if needed, by increasing further the train set size. 

\subsection{Sampling and thermodynamic integration}

To compute finite temperature properties we perform different kinds of standard and accelerated molecular dynamics simulations. Unless otherwise specified, all simulations use a timestep of 2 fs, with a BAOAB integrator\cite{leim+13jcp}. Efficient constant-temperature sampling is achieved by combining stochastic velocity rescaling\cite{buss+07jcp} and a colored-noise langevin thermostat\cite{ceri+10jctc}, as implemented in i-PI~\cite{kapi+19cpc}. Energies and forces are computed using the n2p2~\cite{sing+19jctc} package interfaced with LAMMPS~\cite{plim95jcp}.
In constant-pressure simulations, the pressure is controlled with the Bussi-Zykova-Parrinello barostat \cite{buss+09jcp,ceri+14cpc}. The friction parameters of the barostat and the thermostat are set to 225 fs and 100 fs respectively. 
To compute self-diffusion coefficients and viscosity we applied weak global velocity rescaling thermostat~\cite{buss+07jcp} with a 1 ps time constant, which improves statistical sampling without affecting dynamical properties. 
To shrink the statistical error on computing the bulk modulus, the heat capacity and the stability of defects, we run replica exchange molecular dynamics (REMD) \cite{Sugita1999,Okabe2001,Petraglia_2015} with a exchange time of 40 fs.
Examples of simulations, and the complete set of parameters chosen for interface pinning and metadynamics simulations is provided as commented input files in the SI.

\section{Results}\label{sec:results}
 
After having discussed the construction of the machine-learning models we use, and the details of the reference calculations, we now present results that can be obtained when applying them to the prediction of the atomic-scale properties of elemental Ni. We first validate the model by comparing its predictions with explicit density-functional calculations, and then proceed to compute a large number of finite-temperature properties, for which we compare with experimental data and/or previous literature results. We also use an EAM potential\cite{pun_development_2009} to gauge the typical accuracy of a well-established empirical model, and to contrast it with that of a DFT-trained ML scheme. Whenever we compare two computational schemes, we use exactly the same simulation protocol, to ensure that any discrepancy is due to the potential energy surface, and not to finite size effects or other simulation details. 

\subsection{\label{subsec:valid}Validation of the NN potential}

To provide a first benchmark of the accuracy of the NNP we predict a few simple, static-lattice properties that can be readily recomputed by DFT. We present bulk properties, defects and interfacial energetics. Most of these quantities are explicitly associated with structures that are included in the training set. For this reason, these tests serve more to demonstrate how the training error is reflected on the properties of interest, rather than to assess the transferability of the NN. 

\begin{table}[bthp]
\centering
\begin{tabular}{l | c c c | c c c c }
\hline
\hline
 & \multicolumn{3}{c|}{$\Delta E|_{fcc}$/(meV/at.)}& \multicolumn{4}{c}{$a_0$/\AA} \\
 & NNP  & DFT  & EAM & NNP  & DFT  & EAM  & Exp. \\ \hline
\emph{fcc}   & - & - & - & 3.5168 & 3.5175 &  3.5200 & 3.524 \\
\emph{hcp}               & 20.8 & 21.3 & 22.2 & 2.4873 & 2.4801& 2.4819  \\ 
\emph{}\ \ ($c_0$)               &  &  &  & 4.0829 & 4.0971& 4.1048   \\ 
\emph{bcc}              & 98.3 & 98.0 & 67.4  & 2.7968& 2.7962& 2.7687     \\ \hline
\end{tabular}
\caption{The atomic bulk energies of hcp and bcc ideal crystalline structures with respect to the fcc bulk equilibrated at 0K, as well as the equilibrium lattice parameters. Experiments are taken from \cite{swartzendruber1991fe} where the measurements were carried out at 20$^\circ$C. }
\label{tab:bulk-energies}
\end{table}

\subsubsection{Structure and stability of fcc, hcp and bcc phases}

The stable structure for crystalline nickel at room temperature and pressure is \emph{fcc}. Higher-energy, meta-stable phases, however, can play a role in different portions of the phase diagram, in the presence of defects, or just to increase the transferability of the NNP. 
Table~\ref{tab:bulk-energies} shows the 0K lattice energy of \emph{bcc} and \emph{hcp} configurations relative to the \emph{fcc} ground state, as well as the relaxed lattice parameters. The sub-meV accuracy of the NN is consistent with the overall test and train set errors; the large discrepancy observed for the EAM model for the \emph{bcc} phase is unsurprising, given that the empirical potential is optimized for the stable phases of Ni. Lattice parameters are in excellent agreement with the DFT reference values.

\subsubsection{Elastic constants and bulk modulus}

The bulk modulus and the elastic constants characterise the response of a material to isotropic and anisotropic deformations. 
Together with structural properties such as the zero-temperature lattice constants they can be easily measured experimentally and do not require substantial computational resources to obtain from electronic structure calculations, making them good references for benchmarking. 
We compute the bulk modulus of \emph{fcc} nickel and its derivative by evaluating the change in potential energy when introducing finite isotropic deformations (up to 5\% of the equilibrium lattice parameter), and fitting the resulting energy-volume curve to a Birch-Murnaghan equation \cite{birch1947finite}: 
\begin{equation}
\begin{split}
E(V)=E_{0}+\frac{9 V_{0} B_{0}}{16}\Biggl\{\left[\left(\frac{V_{0}}{V}\right)^{\frac{2}{3}}-1\right]^{3} B_{0}^{\prime} +\\+\left[\left(\frac{V_{0}}{V}\right)^{\frac{2}{3}}-1\right]^{2} \left[6-4\left(\frac{V_{0}}{V}\right)^{\frac{2}{3}}\right] \Biggr\}
\end{split}
\label{eq:bir-mur}
\end{equation}
where $E_0$ is the minimum lattice energy, $V_0$ is the reference volume, $B_0$ is the bulk modulus, and $B^{\prime}_0$ is the derivative of the bulk modulus with respect to pressure.

\begin{table}[btp]
\centering
\begin{tabular}{l l l l l}
\hline
\hline
                  & NNP & EAM & DFT & Exp. \\ \hline
$B$/GPa & 204 & 180 & 205 & 183        \\ 
$B'$ & 4.3 & 4.6 & 4.7 &  --        \\ 
$C_{11}$/GPa     & 275 & 236 & 277 & 243    \\ 
$C_{12}$/GPa     & 167 & 154 & 169 & 153    \\ 
$C_{44}$/GPa     & 130 & 127 & 133 & 128    \\ \hline

\end{tabular}
\caption{Bulk modulus, bulk modulus derivative $B'$ and elastic constants for the current NNP, EAM potential, DFT and experimental results from Ref~\citenum{zhang2001high}.}
\label{tabl:elastic}
\end{table}

For a cubic material the bulk modulus is also linked to the second order elastic constants by the expression:
\begin{equation}
    B = \frac{1}{3}(C_{11}+2C_{12})
\end{equation}
where the standard Voigt notation is being used for the indices.  
We estimate the elastic constants by examining the strain energy density for orthorhombic and monoclinic deformations which corresponds to strain tensors of the form:
\begin{subequations}
\begin{align}
E_{orth}(\delta)/V  = \left \{
  \begin{tabular}{ccc}
  $\delta$ & 0 & 0 \\
  0 & $-\delta$ & 0 \\
  0 & 0 & ${\frac{\delta^2}{1-\delta^2}}$ 
  \end{tabular}
\right \} \\
E_{mon}(\delta)/V = \left \{
  \begin{tabular}{ccc}
  0 & $\frac{1}{2}\delta$ & 0 \\
  $\frac{1}{2}\delta$ &0 & 0 \\
  0 & 0 & ${\frac{\delta^2}{4-\delta^2}}$ 
  \end{tabular}
\right \}
\end{align}
\label{eq:orthmon}
\end{subequations}
Both matrices define deformations which preserve the volume $V$ of the examined system. 
The corresponding strain energy densities $\Delta E_{orth}(\delta)/V$ and $\Delta E_{mon}(\delta)/V$ are given by:
\begin{subequations}
\begin{align}
\Delta E_{orth}(\delta)/V = (E_{tot}(\delta)- E_0)/V = (C_{11} - C_{12})\delta^2 + \mathcal{O}  (\delta^3)
 \\
\Delta E_{mon}(\delta)/V = (E_{tot}(\delta) - E_0)/V = \frac{1}{2}C_{44}\delta^2 + \mathcal{O}  (\delta^3),
\end{align}
\label{eq:elastic-cnst}
\end{subequations}
where $E_{tot}(\delta)$ denotes the total energy of the deformed system, $E_0$ is the ideal bulk energy or $E_{tot}(\delta = 0)$. 
We compute energies for values of $\delta \leq 10\%$, and estimate the elastic constants by fitting the resulting curves to Eq.~\eqref{eq:elastic-cnst}. Results, shown in Table~\ref{tabl:elastic}, indicate that the NN reproduces the DFT elastic constants with high accuracy (an error around 2\%{}), and is consistent with previous results for single element bulk metals ~\cite{drag+18prm, koba+17prm, szla+14prb} which also report an error smaller than 4\%{} between DFT and machine-learning potentials. %

\begin{figure}
    \includegraphics[width=1.05\columnwidth]{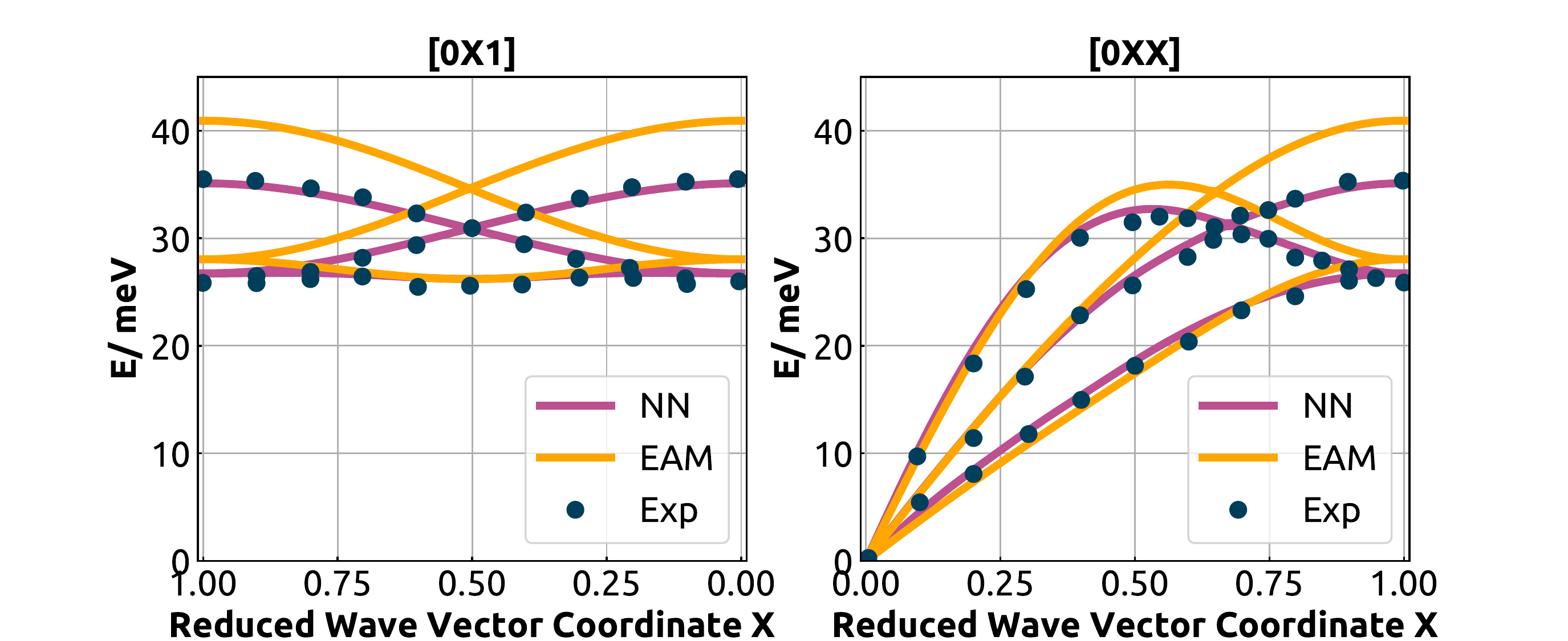}
    \caption{Phonon dispersion curves for the EAM potential (yellow), the NNP potential (purple) and experiment(blue dots)}
    \label{fig:phonon}
\end{figure}

\subsubsection{Phonons}

Phonon dispersion curves describe the elastic response of the interatomic potential to a plane wave deformation of wavevector $\mathbf{q}$, and can measured by inelastic neutron or X-ray scattering. DFT has been shown to reproduce closely experimental phonon curves for pure Ni\cite{wang_thermodynamic_2004}. For this reason, we compare the NNP and the EAM potential with the experimental results.

The phonon dispersion curves for NNP and EAM potential have been obtained with the small displacement method as implemented in the PHON package \cite{ase-paper, wang2010mixed, Alf2009}. In the frame of this method, the position of each atom in the primitive cell is slightly distorted. 
The force constant matrix is constructed by computing forces acting on all the other atoms in the crystal, using the DFT equilibrium volume. This force constant matrix is used to compute the dynamical matrix at any chosen $\mathbf{q}$-vector in the Brillouin zone, which is then diagonalized to yield the squares of the phonon frequencies. 
The resulting dispersion curves are shown in Fig.~\ref{fig:phonon}. NNP results are in excellent agreement with experiments, while those obtained with the EAM show a deviation up to 20\%{} for the longitudinal mode at the brillouin-zone edge.

\subsubsection{Formation energies of point defects}

At finite temperature any crystalline system contains an equilibrium concentration of point defects, such as vacancies and interstitial atoms.  
The ab initio calculation of the single point defect formation energies can be achieved  with low effort from the expression:
\begin{equation}
    E_{def}^{f}=E_{def}(N_{def})-\left[N_{def} / N_{0}\right] E_{0} \label{eq:ef-def}
\end{equation}
where $E_{def}$ is the final energy of the system with a defect after full ionic relaxation, $N_{def}$ -- number of atoms in the system with a defect, while $N_0$ and $E_0$ indicate the number of atoms and the energy of a reference supercell corresponding to ideal crystal.

\begin{table}[bthp]
\centering
\begin{tabular}{l l l l l}
\hline
\hline
                  & NNP  & EAM  & DFT  & Experiment\\\hline
$E^{f}_{vac}$, eV & 1.52 & 1.57 & 1.51 & 1.4(900-1400K)        \\ 
$E^{f}_{int}$, eV & 4.17 & 4.01 & 4.2  &            \\ \hline
\end{tabular}
\caption{Formation energies of single vacancy and interstitial in bulk Ni for NNP, EAM, DFT and experiment\cite{glazkov1987formation} .}
\label{tabl:spd}
\end{table}
We use a relatively large cell size ($3\times 3\times 3$ conventional unit cells, corresponding to 108 atoms) which ensures  that the interaction of defects through periodic boundaries is negligible. Ionic positions have been fully relaxed using the BFGS algorithm~\cite{broyden1970convergence,fletcher1970new,goldfarb1970family,shanno1970conditioning}. As shown in Table~\ref{tabl:spd}, the NNP is in excellent agreement with reference DFT calculations, and in semi-quantitative agreement with experimental data\cite{glazkov1987formation}, which is however collected at finite temperature, the effect of which is discussed in Section~\ref{sub:tdep-defects}.  %

\subsubsection{Generalised stacking fault}

\begin{figure}[tbp]
    \includegraphics[width=1.0\columnwidth]{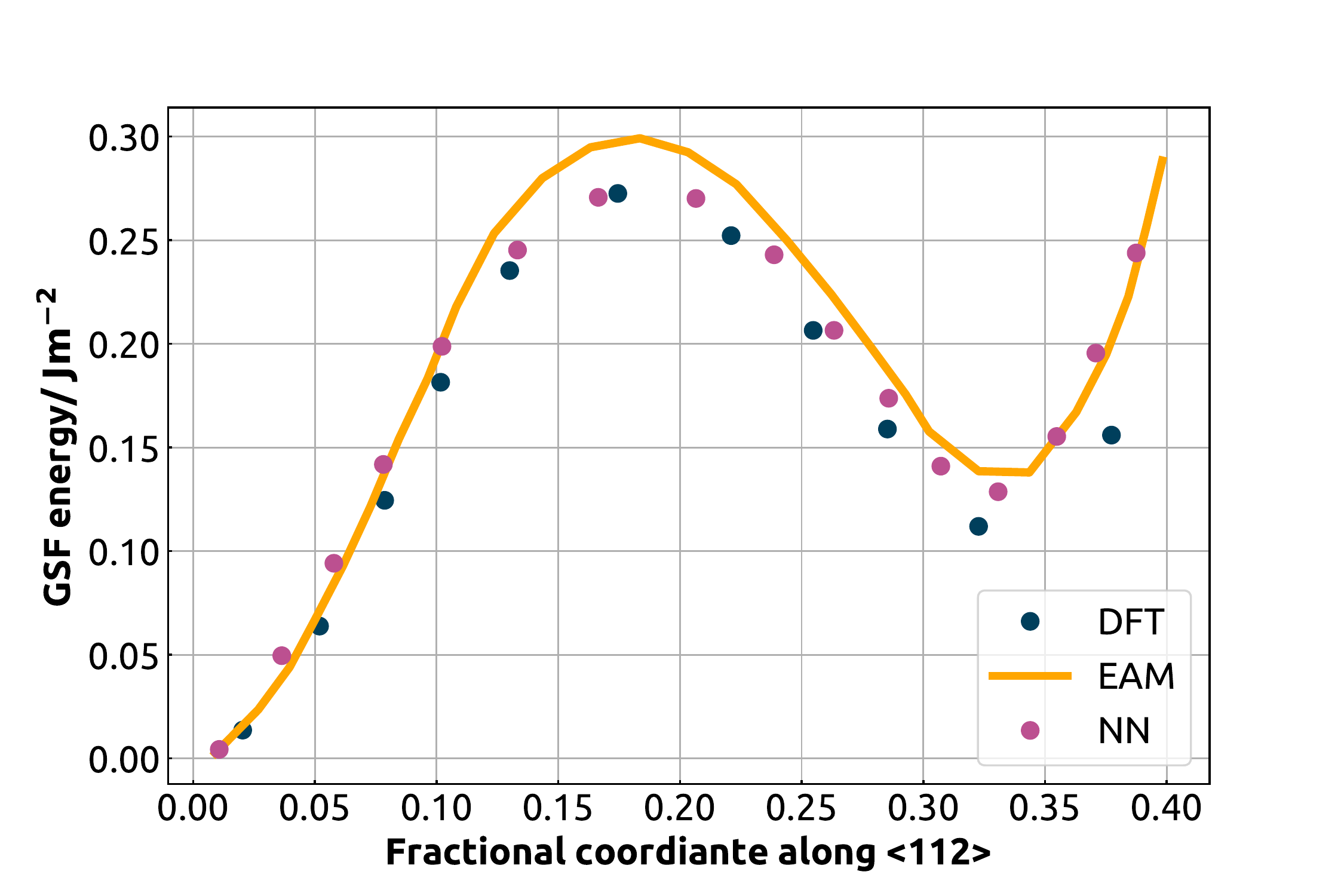}
    \caption{Generalized stacking fault curve for bulk Ni along the [112] direction computed using DFT (blue dots), EAM potential (yellow curve), and the present NN potential (purple dots).}
    \label{fig:gsf}
\end{figure}

The Generalised stacking fault (GSF) energy is an important property that is related to the response of a material to plastic deformation and fracture. The GSF reports on the energy cost associated with the slip of the crystal along a plane of atoms, with the geometric nature of the deformation being determined by the crystal lattice and symmetries.
The only point along a GSF curve that can be probed experimentally is the one corresponding to an intrinsic stacking fault geometry. However it is possible to compute the full curve in simulations, by tilting the repeat vector of an ideal crystalline lattice in a slip plane while keeping all the atoms fixed.\cite{yin_comprehensive_2017} The shift of PBCs creates a stacking fault. The deformed system is then relaxed along the direction orthogonal to the slip plane. The full GSF curve can be sampled by introducing larger and larger tilt angles. The GSF energy is defined as:
\begin{equation}
    \gamma^{SF}(x,y) = \frac{E[N](x,y) - [N/N_0]E_0}{A_{xy}},
\end{equation}
where $A_{xy}$ is the cross-section of the supercell.
For reference DFT calculations we used an elongated supercell, with a 1x1 dimension along the fixed in-plane lattice vectors, and a 4-fold replication along the [111] direction to minimize interactions between the periodic images of the SF.
The NNP reproduces accurately points computed with DFT, while the EAM potential slightly overestimates stable and unstable stacking fault energies (Fig. \ref{fig:gsf}).

\begin{figure}[tbp]
    \includegraphics[width=1.0\columnwidth]{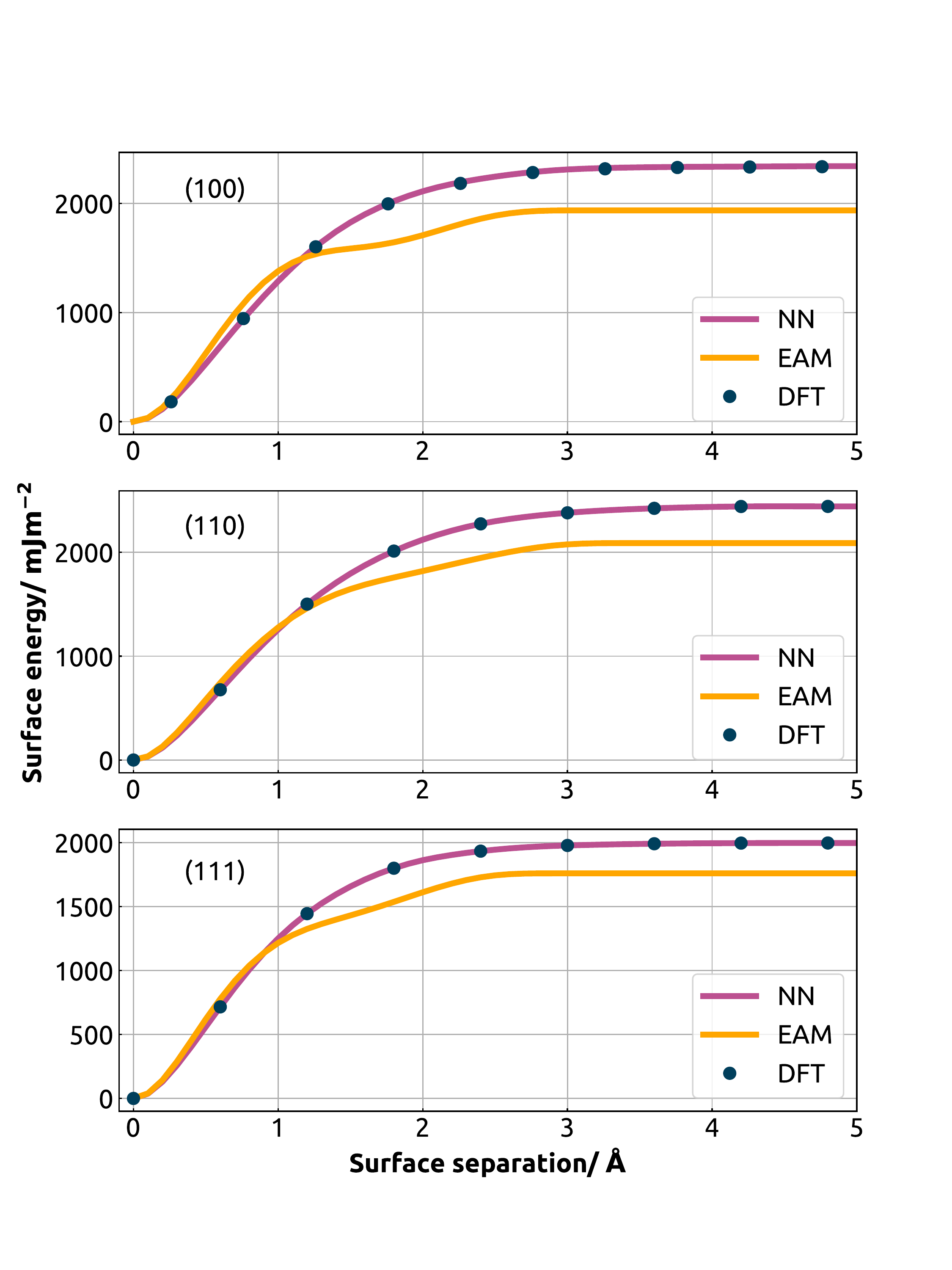}
    \caption{The energy  vs rigid separation across different surface orientations for DFT (blue dots), EAM (yellow curves) and NNP (purple curves).  }
    \label{fig:clen}
\end{figure}

\subsubsection{Rigid surface separation}

The surface energy of solids controls many technologically-relevant phenomena such as fracture, nucleation, morphological surface properties etc. 
Experimentally this property is affected by the presence of defects and impurities, and by surface reconstruction. Computationally, a rigid cleaving of the ideal bulk makes it possible to easily determine whether a potential provides a satisfactory description of the formation of a free surface. 

\begin{table}[tbp]
\centering
\begin{tabular}{lllll}
\hline
\hline
Surfaces, $mJ/m^2$ & NNP  & EAM  & DFT  & Experiment \\ \hline
(110)              & 2468 & 2087 & 2440 & 2280       \\ 
(001)              & 2351 & 1936 & 2337 & 2280       \\ 
(111)              & 2004 & 1759 & 1995 & 2280       \\ \hline
\end{tabular}
\caption{The surface energy of different surface orientations for NNP, EAM, DFT and experiment. The experimental value is averaged over orientations.\cite{murr1975interfacial} }
\label{tabl:cep}
\end{table}

The cleaving potential is computed by evaluating the energy of a bulk solid configuration, in which the lattice spacing between two planes is artificially increased by a separation $d$. 
Given the energy $E(N,d)$ of a supercell with $N$ atoms and cross-section $A_{xy}$, the rigid-surface cleaving potential is defined as 
\begin{equation}
    \gamma^{surf}(d) = \frac{E(N,d) - [N/N_0]E_0}{2A_{xy}}
\label{eq:solsurf}
\end{equation}
where $E_0$ is the energy of a reference bulk configuration with $N_0$ atoms. 
For our reference calculations we consider supercells elongated along 
the (111), (001), and (110) directions, with 8 atomic layers in the direction orthogonal to the surface. %
The EAM potential captures correctly the order of surfaces stability (table \ref{tabl:cep}), although with poor quantitative agreement with DFT, which matches well the experimental estimate\cite{murr1975interfacial} (which is an average over multiple orientations). Similar to what was observed  for Al in Ref.~ \citenum{kobayashi_neural_2017}, the EAM cleaving potential displays an unphysical step-like behavior.

\subsection{\label{subsec:fprop}Finite-temperature properties}

Benchmarks on static lattice calculations, such as those discussed in the previous Section, give confidence on the accuracy of the MLP, as they can be compared with little effort with reference DFT calculations. 
This Section, instead, focuses on properties that require the evaluation of thermodynamic averages at finite temperature. In the low-$T$ regime, quantum fluctuations of the nuclei are also important, while at high temperature magnetic and electronic excitations also play a role in determining the thermophysical properties of Ni. 
Given that most of the simulations we report in this Section would be impractical when coupled to explicit quantum calculations, we cannot directly compare our results to the DFT reference. We do however compare with existing force fields and with experiments, even though we cannot disentangle the errors associated with the underlying electronic-structure approximations, and those stemming from the NN fit. 

\begin{figure}
    \centering
    \begin{subfigure}[t]{1.\columnwidth}
        \centering
        \includegraphics[width=1.\columnwidth]{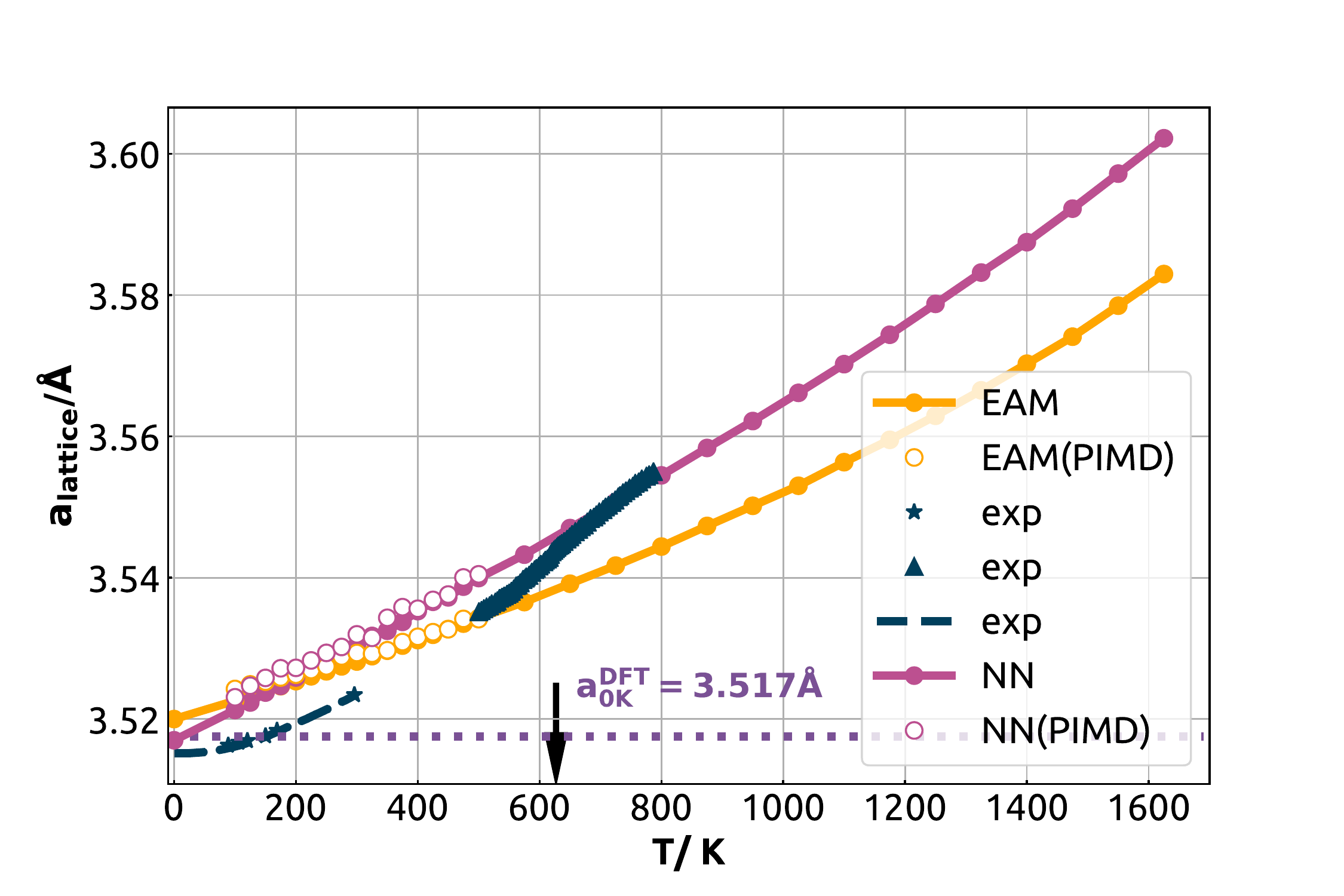}
        \caption{Lattice parameter of Ni as a function of temperature for EAM and NN computed classically (filled circles) and with PIMD (empty circles). Experimental data is presented by triangles\cite{Yousuf_1986}, starts and  dashed curve \cite{Bandyopadhyay1977}.} Black arrow points at the Curie point for Ni. 
     \label{fig:alat}
    \end{subfigure}
    \begin{subfigure}[t]{1.\columnwidth}
        \centering
        \includegraphics[width=1\columnwidth]{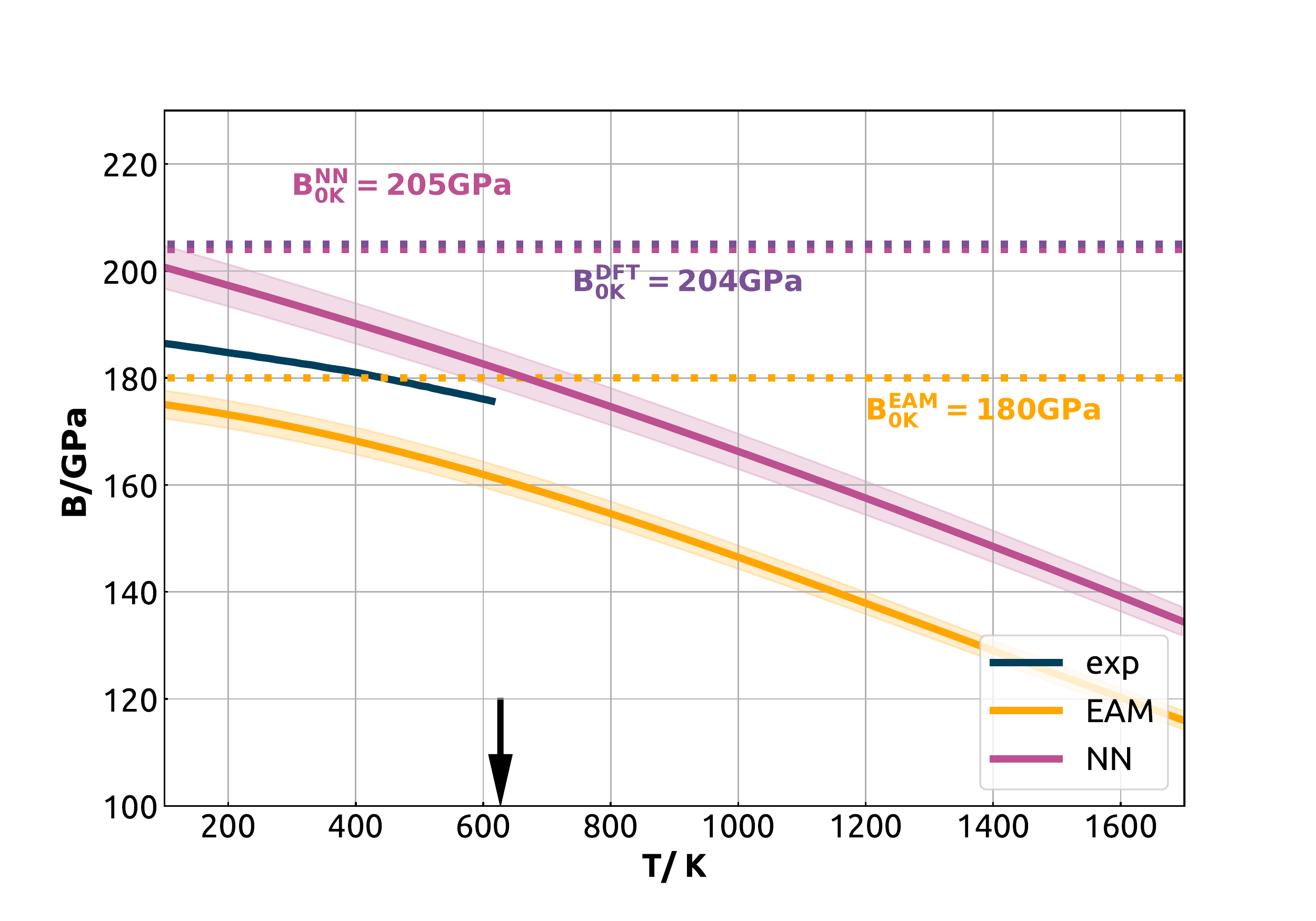}
        \caption{Bulk modulus of Ni as a function of temperature for EAM and NN. Shaded areas represent corresponding statistical errors. Blue curve indicates experimental data from \citenum{Shimizu_1978}. Black arrow points at the Curie point for Ni.}
    \label{fig:bm}
    \end{subfigure}
\caption{}
\end{figure}

\subsubsection{Structural and elastic properties at finite temperatures}
We begin by revisiting the bulk properties of Ni incorporating the effect of fluctuations. The Debye temperature of Nickel is around 400~K, and so one can expect a significant effect associated with quantum fluctuations of the nuclei up to and above room temperature. 
For this reason, we perform simulations using both classical molecular dynamics (that are valid in the high-temperature limit) and with path integral molecular dynamics (PIMD)\cite{parr-rahm84jcp,feyn-hibb65book,tuck08book} (that incorporate nuclear quantum effects in the low temperature limit). To accelerate convergence of PIMD simulations, we use a finite-difference integrator~\cite{kapi+16jcp2} for the fourth-order Suzuki-Chin factorization of the path integral partition function,~\cite{suzu95pla,chin97pla} as implemented in i-PI\cite{kapi+19cpc}, that yields converged observables down to about 100K with only four replicas.

The top panel of Fig.~\ref{fig:bm} shows the behavior of the lattice parameter with temperature, as obtained from REMD simulations of a box of 108 atoms, run for approximately 150ps at each temperature with a possibility to swap between replicas every 40fs. The thermal expansion is similar between the NN and EAM simulations, and both are in good agreement with experiments~\cite{Yousuf_1986,Bandyopadhyay1977}. 
Both the EAM and the NN cannot capture the effects of the ferromagnetic transition: the EAM is fitted to low-temperature structural parameters and underestimates the lattice parameter in the high-$T$ regime, while the NN, that is fitted to a non-polarized DFT reference, shows a better agreement above the Curie temperature, and a overestimates the lattice parameter in the ferromagnetic phase. 
Quantum effects on the lattice parameters are small even below the Debye temperature, which justifies using a classical expression to estimate the bulk modulus in this temperature range by considering the volume fluctuations at constant pressure: 
\begin{equation}
    B(T) = \frac{\langle V\rangle k_B T}{\left\langle V^{2}\right\rangle-\langle V\rangle^{2}}.
\end{equation}
As shown in Fig.~\ref{fig:bm}, the bulk modulus shows a substantial dependency on temperature, with EAM and NN bracketing experimental observations, and exhibiting a similar trend up to the melting point. 

\begin{figure}
    \includegraphics[width=1.\columnwidth]{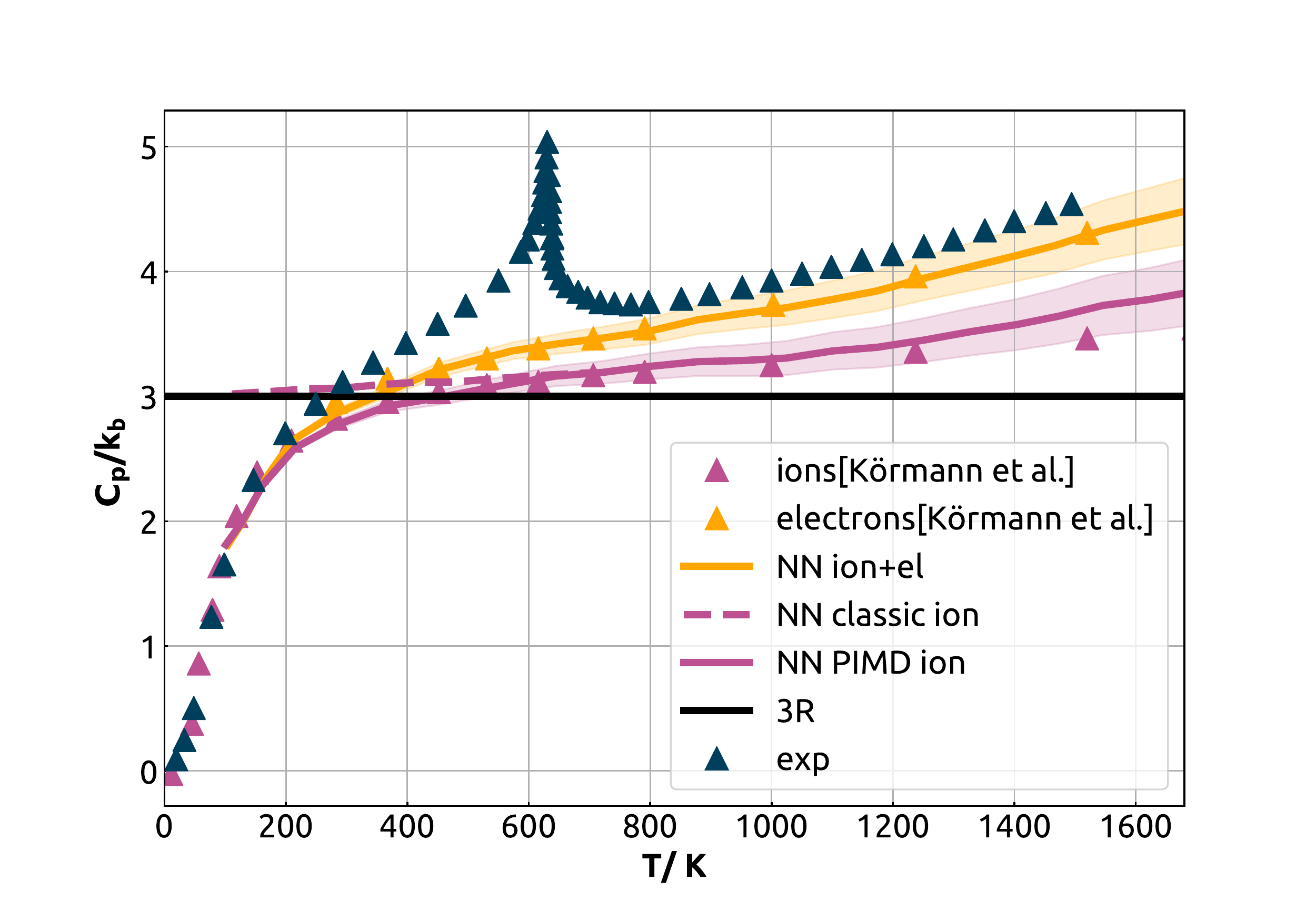}
    \caption{Constant pressure heat capacity $C_p$ as a function of temperature. Triangles indicate experimental observations, as well as electronic and vibrational contributions computed within quasiharmonic approximation at low-temperature, and by classical ab initio MD at high temperature~\cite{kormann2011role}. Solid lines represent the heat capacity computed in this article with PIMD, and including electronic corrections based on a ML model of the DOS. }
    \label{fig:cp}
\end{figure}

\newcommand{\cp}{C_p}
\subsubsection{Heat capacity}

The constant-pressure heat capacity $\cp$ of a ferromagnetic metal such as Nickel is a very challenging quantity for modeling, because it contains features that are associated with excitations on different degrees of freedom and energy scales~\cite{drag+18prm}. As shown in Fig.~\ref{fig:cp}, the experimental curve shows a low-temperature limit which is dominated by quantum nuclear effects, tending to zero at low temperature, a peak around the Curie temperature, associated with the ferromagnetic phase transition, and a pronounced increase above the Dulong-Petit limit at high temperature, that is linked to electronic excitations. 
Thus, a very accurate interatomic potential is not sufficient to accurately predict the full $\cp$ curve. 
Within the adiabatic approximation, ionic, electronic and magnetic contributions to the heat capacity could be described separately, provided one can treat them explicitly, as one would do in ab initio molecular dynamics.
Here we present a first application of an integrated ML model that incorporates properties beyond the interatomic potential, to have access to contributions beyond those controlled by ionic fluctuations. 
We focus in particular on the electronic effects, that can be estimated, within a rigid band approximation, from the knowledge of the electron density of states (DOS). The contribution to the internal energy associated with electronic excitations can be computed as:
\begin{equation}
    U_{\mathrm{DOS}}^{\mathrm{el}}(T)=\int_{-\infty}^{\infty} \varepsilon  D(\varepsilon) f( \varepsilon-\varepsilon_F,T) \mathrm{d} \varepsilon-\int_{-\infty}^{\varepsilon_{\mathrm{F}}} \varepsilon D(\varepsilon)  \mathrm{d} \varepsilon
\end{equation}
where $D(\varepsilon)$ represents the averaged DOS over the trajectory at a fixed temperature, $f( \varepsilon-\varepsilon_F,T)$ is the Fermi function evaluated at temperature $T$, and the Fermi energy $\varepsilon_F$ is determined separately in the two integrals by enforcing charge neutrality.
We used a recently-introduced machine learning model of the DOS~\cite{benmahmoud2020arxiv}, trained as discussed in Section~\ref{sec:dos-theory},  to predict the electronic density of states (DOS) for every frame of the REMD simulation, which was then used to estimate the electronic energy $U_\text{DOS}$ and, by finite differences, the electronic contribution to $\cp$.

In Figure~\ref{fig:cp} we show the heat capacity as a function of temperature computed from classical molecular dynamics (purple dashed line) using the fluctuation formula
\begin{equation}
\cp = \frac{\left<H^2\right>-\left<H\right>^2}{k_B T^2}
\end{equation} 
that deviates dramatically from the experimental curve at low temperature.
Results from PIMD, that are evaluated with a fourth order double virial operator heat capacity estimator\cite{yama05jcp},  (purple solid line) display the correct low-temperature behavior, but underestimate by $\approx$20\%{} the experimental observations at high temperature. The discrepancy is due to electronic contributions, and indeed the curve that incorporates these using the ML model of the DOS (yellow solid line) are in almost perfect agreement with high-temperature measurements, and with previous results obtained, with heroic efforts, using density functional theory and quasi-harmonic simulations in the low-temperature regime~\cite{kormann2011role}. 
Incorporating quantum nuclei and electronic fluctuations lead to remarkably good agreement with experiments, except for the region around the Curie temperature, where magnetic excitations become important. Even though we do not incorporate them in this model, adding a description of magnetism constitutes an interesting direction for future studies. 

\begin{figure}
    \includegraphics[width=1.\columnwidth]{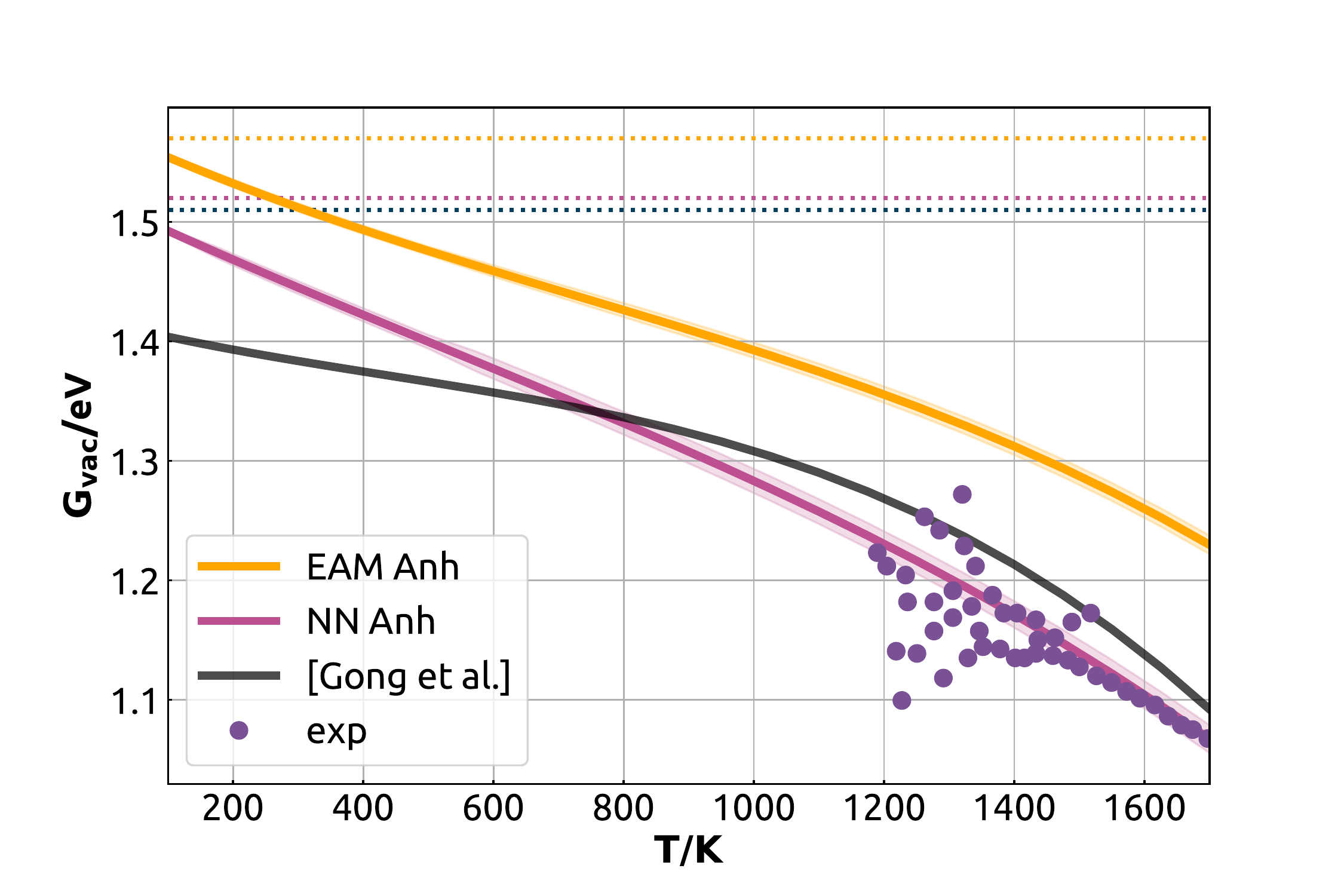}
    \caption{The Gibbs free energy of a single vacancy in fcc obtained with thermodynamic integration relative to the potential energy difference at 0 K for EAM and NN compared to the curve reported in \citenum{gong_temperature_2018} based on \emph{ab initio} calculations, and available experimental data\cite{Metsue_2014,wycisk1978quenching,scholz2001messungen, michot1977influence}. Dashed lines indicate the level of 0K energy of formation for NN, EAM and DFT calculated in this work.}
    \label{fig:ti}
\end{figure}

\subsubsection{Stability of defects}
\label{sub:tdep-defects}

Finite-temperature and quantum fluctuations also affect the stability of defects. 
We estimate their contribution using thermodynamic integration (TI)~\cite{tuck08book,ghir+15prl,chen-ceri18prb} that makes it possible to estimate the absolute free energy of a thermodynamic state by a sequence of transformations, and use the values for two different states to estimate their relative stability.
For instance, the Gibbs free energy of a single point defect can be easily found with an expression analogous to Eq.~\eqref{eq:ef-def}:
\begin{equation}
    G_{\mathrm{d}}=G_{\mathrm{defect}}-\frac{N_{\mathrm{defect}}}{N_{\mathrm{perfect}}}G_{\mathrm{perfect}},
\end{equation}
where $G_{\mathrm{defect}}$ and $G_{\mathrm{perfect}}$ refer to the absolute free energies of two supercells, one of which includes the defect. 

Here we use the free energy of the harmonic crystal as the reference state, which can be straightforwardly computed as:
\begin{equation}
    F_{\mathrm{h}}\left(V, T_{0}\right)=k_{B} T_{0} \sum_{i=1}^{3 N-3} \ln \frac{\hbar \omega_{i}}{k_{B} T_{0}}
\end{equation}
where $\omega_{i}$ are phonon frequencies of the crystal with $N$ atoms,  and $T_0$ a low temperature chosen so that the system is close to a local minimum of the potential energy. 
Note that we use the classical expression, because we are ultimately interested in high-temperature values of the free energy. If one wanted to estimate the anharmonic free energy at low temperature, it is possible to do so by a further thermodynamic integration step~\cite{habe-mano11jcp,ross+16prl,chen+18prl}.
Starting from the harmonic reference, one then performs the actual TI step, that involves  parameterising a Hamiltonian $\mathcal{H}(\lambda)$ in such a way that $\mathcal{H}(\lambda = 0)$ corresponds to the harmonic potential and $\mathcal{H}(\lambda = 1)$ to the real system. One then evaluates numerically the integral
\begin{equation}
 \Delta F=F\left(\lambda = 1\right)-F\left(\lambda = 0\right)=\int_{0}^{1} \mathrm{d} \lambda\left\langle\frac{\partial \mathcal{H}}{\partial \lambda}\right\rangle_{\lambda}
\end{equation}
to give the free energy difference between the systems, which is the anharmonic correction to the free energy.

By choosing a sufficiently low $T_0$, the system is very close to being harmonic, and this term is small and can be computed easily, possibly even just by free energy perturbation. 
In order to convert between constant-volume and constant-pressure boundary conditions, we perform a constant pressure simulation in conditions that give a mean volume close to that used to compute $F$, and evaluate the distribution of volumes $\rho(V|p,T)$. The Gibbs free energy is then given by 
\begin{equation}
    G(N,p,T) = p V + F(N,V,T) + k_b T \ln{\left[\rho(V|p,T)\frac{V}{N}\right]},
    \label{eq:ftog}
\end{equation}
which is based on the definition of the isobaric partition function $\mathcal{Z} = \int \mathrm{d}V e^{-\beta p V} e^{-\beta F(N,V,T)} NV^{-1}$ discussed in Ref.~\citenum{Han_2001}.

To evaluate the Gibbs free energy at higher temperature, one can then perform a series of $NpT$ simulations at different values of $T$ -- possibly using replica exchange to enhance statistical convergence -- and evaluate a TI estimate of 
\begin{equation}
     \frac{G\left(p, T_{1}\right)}{k_{B} T_{1}}=\frac{G\left(p, T_{0}\right)}{k_{B} T_{0}}-\int_{T_{0}}^{T_{1}} \frac{\langle \mathcal{H}+pV\rangle}{k_{B} T^{2}} d T
\end{equation}
where $\mathcal{H}$ denotes the total energy. 
As shown in Fig.~\ref{fig:ti}, at high temperature the contribution from finite-temperature free-energy terms is sizable on the scale of the static defect formation energy (which is around $1.5$~eV for the vacancy, see Table~\ref{tabl:spd}). 
Even though TI makes it possible to compute this correction with ab initio molecular dynamics~\cite{Grabowski_2009,Duff_2015,ross+16prl}, the use of a NN potential reduces the cost dramatically, making it feasible to estimate defect formation free energies for more complex defects and for materials with more diverse chemistry and crystallography. 

\begin{figure}
    \includegraphics[width=1.\columnwidth]{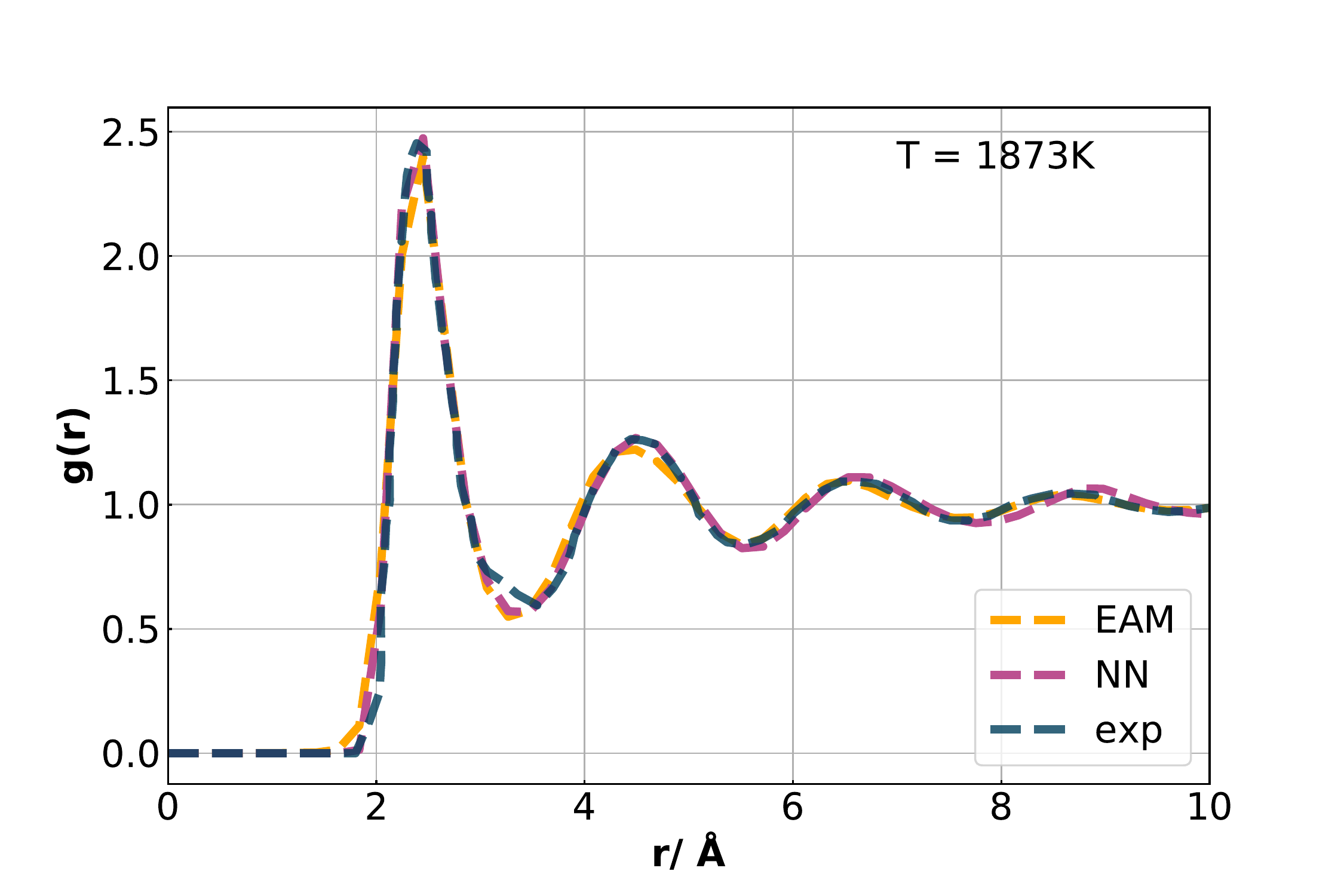}
    \caption{Radial distribution function $g(r)$ of liquid nickel. The experimental curve corresponds to Fourier transform of a structure factor obtained from neutron scattering for liquid nickel at $T = 1873K$ \cite{johnson1976structure}. G(r)'s for EAM and NN models are computed from NVT trajectories at $V|_{P = 0GPa}$ and $T = 1873K$.  }
    \label{fig:rdf}
\end{figure}

\subsubsection{Structure of the melt}

One of the simplest and most direct diagnostics of the accuracy of an interatomic potential in the high-temperature limit involves computing the pair correlation function, $g(r)=\left<\delta(r-\mathbf{r}_{ij})\right>/(4\pi^2r^2\rho)$.
As shown in Fig.~\ref{fig:rdf}, there is an excellent agreement between the NN, the EAM and the  experimental results from neutron scattering data~\cite{johnson1976structure}.
Although the pair correlation function provides only partial information on the structure, the near-perfect agreement indicates that both the EAM and the NN provide an excellent description of the liquid phase of Ni. 

\begin{figure}
    \includegraphics[width=1.0\columnwidth]{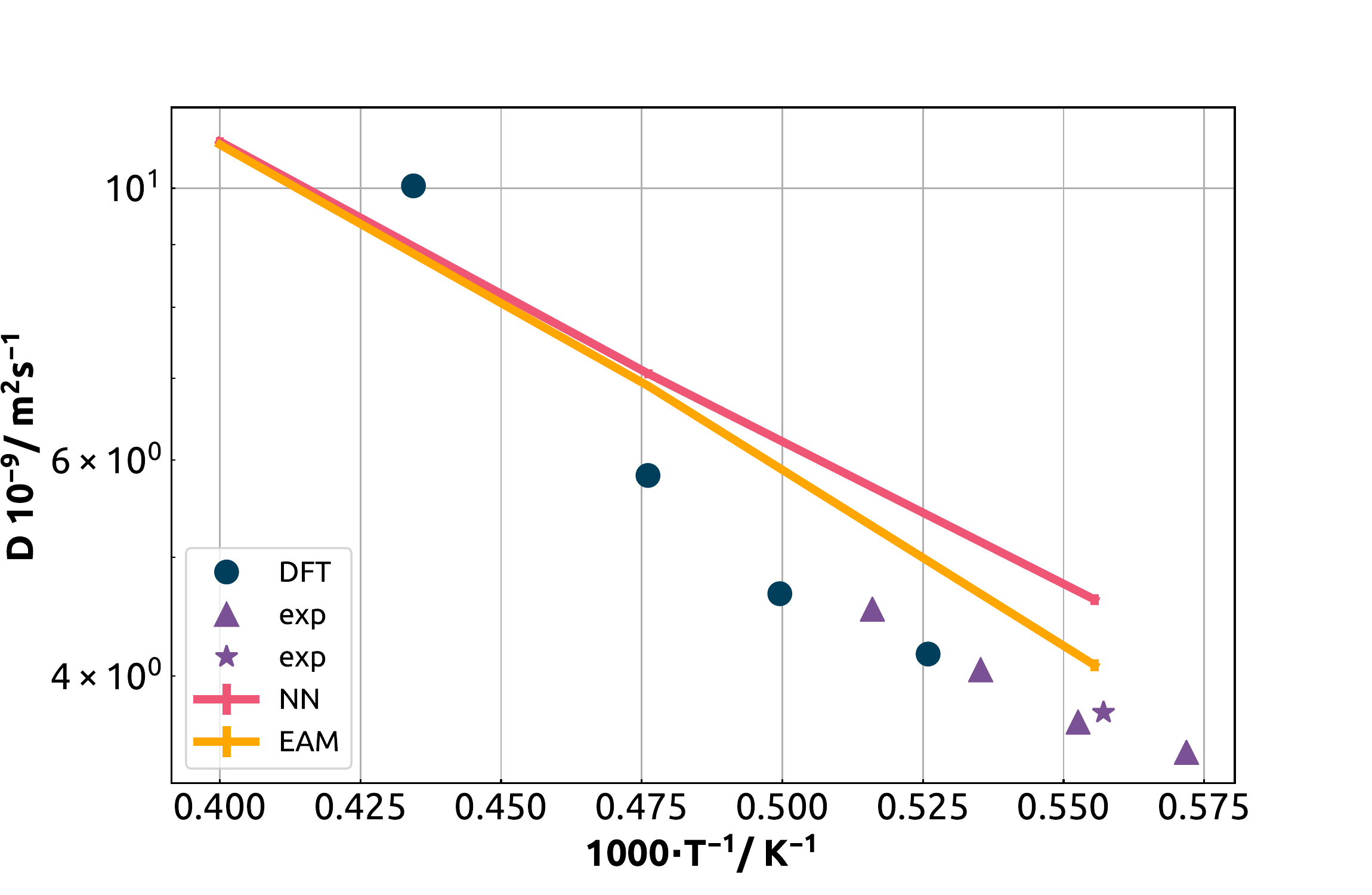}
    \caption{Self-diffusion coefficient of liquid nickel as a function of temperature. The triangles~\cite{meyer2008determination} and the star~\cite{chathoth2004atomic} indicate experimental measurements, dots indicate the result of AIMD simulations reported in Ref.~\citenum{walbruhl2018atomic}.    NNP and EAM results are shown with statistical errorbars.}
    \label{fig:diff}
\end{figure}

\begin{figure}
    \includegraphics[width=1.\columnwidth]{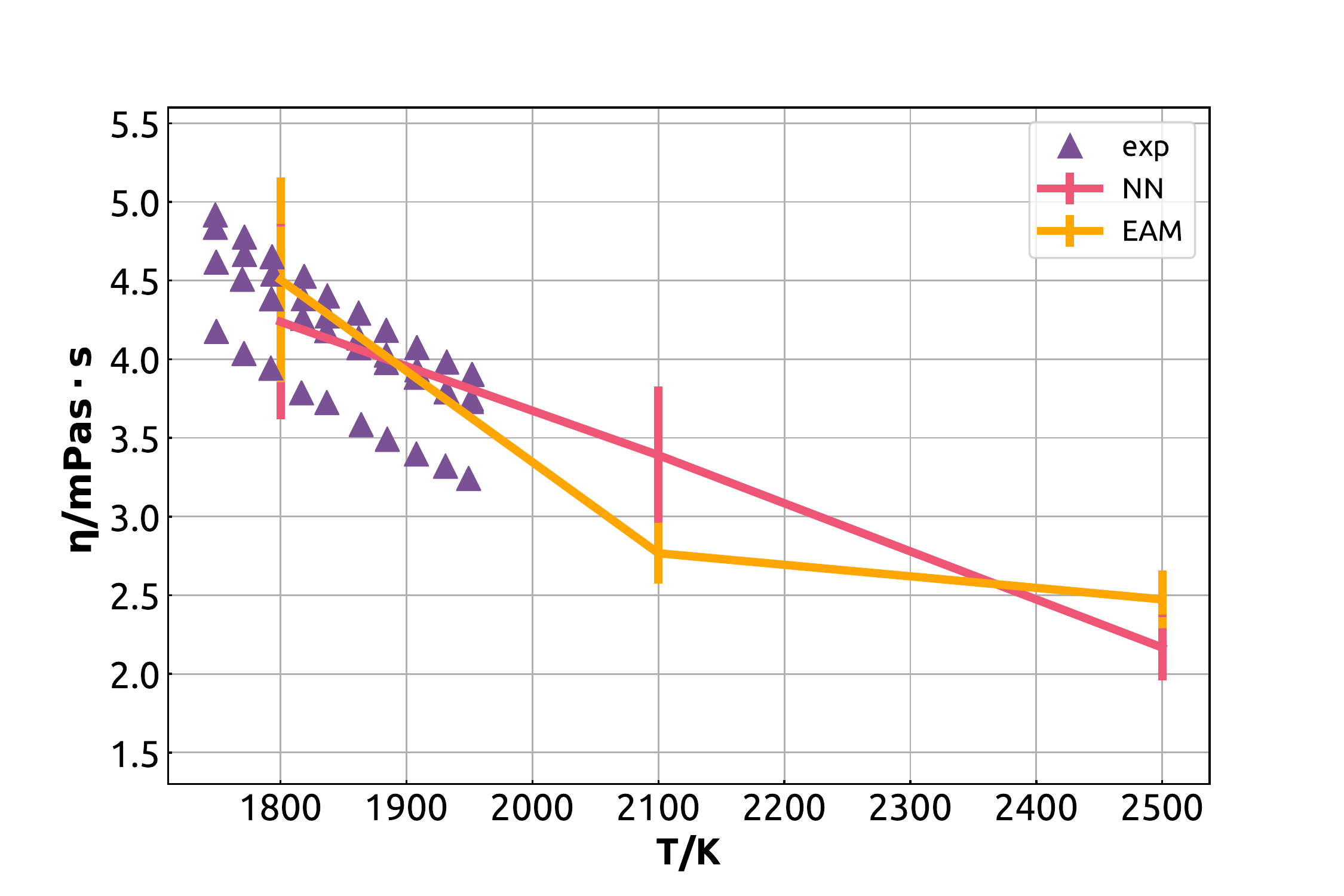}
\caption{Shear viscosity of molten Ni as a function of temperature. The triangles indicate different sets of experiments collected in Ref.\citenum{iida1988physical}, while lines with errorbars correspond to NNP and EAM predictions.}
    \label{fig:viscosity}
\end{figure}

\subsubsection{Self-diffusion coefficients and viscosity}

The self-diffusion coefficient and the viscosity underlie mass transport and convection in the melt. 
They can be computed rather easily from constant-energy (or weakly-thermostatted) molecular dynamics, evaluating the slope of the mean square displacement,
\begin{equation}
D_{\mathrm{sim}}=\lim _{t \rightarrow \infty} \frac{\partial}{\partial t} \frac{\left\langle|\boldsymbol{r}(t)-\boldsymbol{r}(0)|^{2}\right\rangle}{6N}
\end{equation}
that we compute averaging 10 trajectories of 100-100-50(500-500-50)ps each for NN(EAM) simulations involving 108-256-2048 atoms respectively. 
The self-diffusion coefficient has a pronounced dependency on the system size which originates from hydrodynamic self-interaction through the periodic boundary conditions. Thus, comparing the results for a cubic simulation box of length $L$, the diffusion coefficient should be corrected for finite size effects \cite{ D_nweg_1993, Yeh2004}:
\begin{equation}
    D_0 = D_\text{sim} + 2.837297 k_BT/(6\pi\eta L)
\label{eq:diff}
\end{equation}
where $D_{sim}$ is the diffusion coefficient calculated in the simulation, $k_B$ the Boltzmann constant, $T$ the absolute temperature, and $\eta$ the shear viscosity of the liquid. 
Thus, performing simulations at different system size makes it possible to 
extract the viscosity as a fitting parameter of the equation (\ref{eq:diff}) together with $D_0$. 
The diffusion coefficient and viscosity as a function of temperature are shown in Fig.~\ref{fig:diff} and Fig.~\ref{fig:viscosity}, respectively. 
The predicted values for EAM and the NN potential agree with each other, and are in semi-quantitative agreement with experimental measurements.

\begin{figure}[tbp]
    \includegraphics[width=1.0\columnwidth]{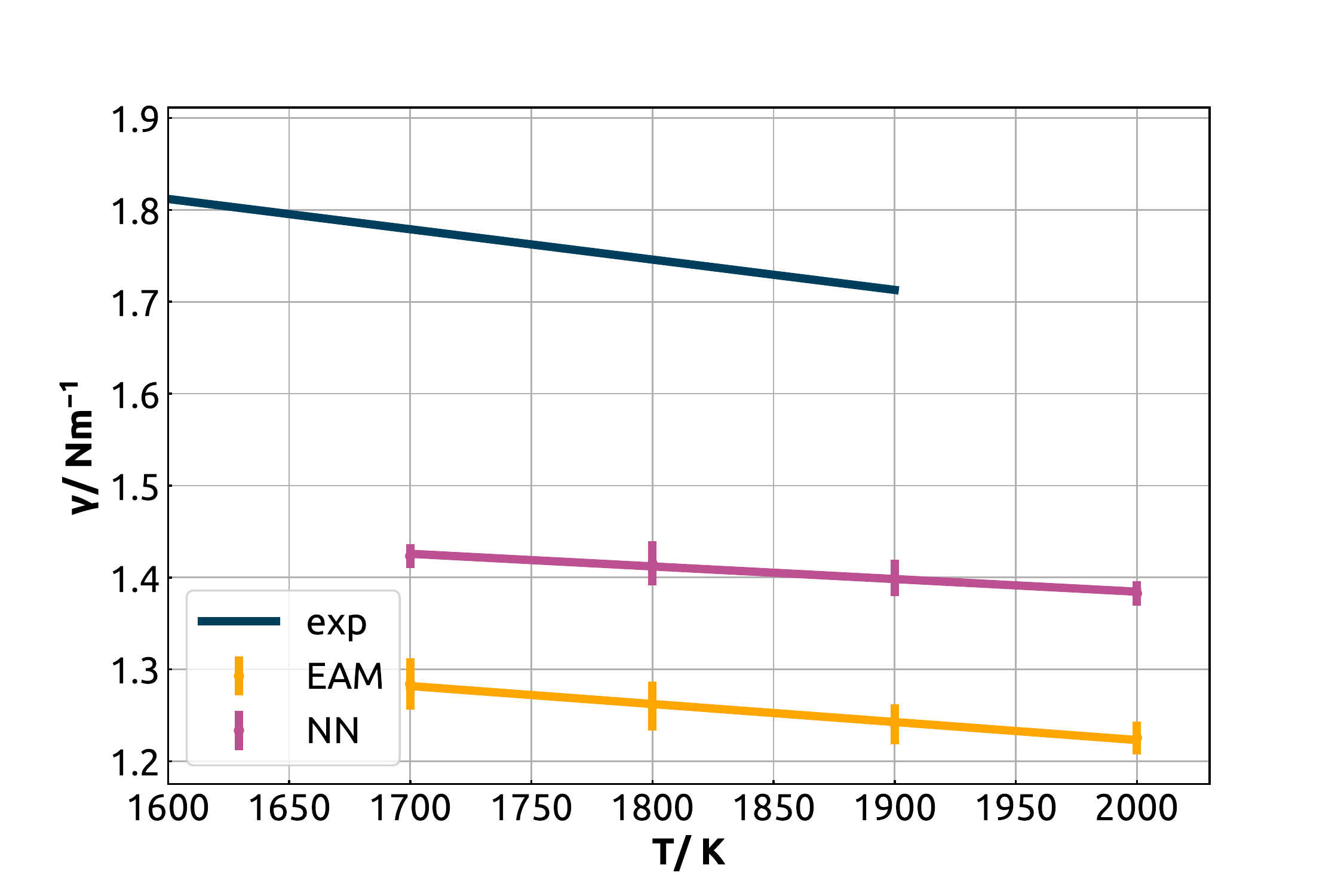}
    \caption{Surface tension of a planar interface as a function of temperature, as computed with NN and an EAM, compared with experimental data from Ref.~\citenum{brillo2005surface}.
    }.
    \label{fig:surftens}
\end{figure}

\newcommand{\glv}{\gamma_{\text{lv}}}
\newcommand{\Tm}{T_{\text{m}}}
\subsubsection{Surface tension}

The liquid-vapor surface energy $\glv$ plays an important role in determining wetting and capillary forces, that are relevant e.g. for additive manufacturing. 
Contrary to solid-vapor surface energies -- that can be reasonably estimated by single-point calculations -- the liquid-vapor surface tension requires averaging over liquid configurations, and simulations size and time scale that are prohibitive for first-principles molecular dynamics. 
A practical simulation protocol involves simulating a planar liquid slab, with two free planar surfaces parallel to xy plane, and computing the integral across the slab of the the normal and tangential components of the stress $\sigma_n$ and $\sigma_t$~\cite{Walton_1983,Cai_2014} 
\begin{equation}
    \glv=\frac{1}{2} \int_{0}^{L_{z}}\left[\sigma_{\mathrm{n}}(z)-\sigma_{\mathrm{t}}(z)\right] \mathrm{d} z
\end{equation}
where $L_z$ is the length of the simulation box. Given the slab geometry, this is equivalent to computing the mean value of the stress of the entire simulation box, using $\sigma_n = \left<\sigma_{zz}\right>$ and $\sigma_t =\left< (\sigma_{xx}+ \sigma_{yy})/2\right>$.
To evaluate $\glv$, we use a slab containing 927 atoms, with a square cross-section of $\sim$
1000\AA$^2$ and $\sim$ 10\AA{} spacing between the surfaces, averaging over 400ps of molecular dynamics simulations.
As shown in Fig.~\ref{fig:surftens}, there is a rather large discrepancy between theoretical and experimental results for the surface tension, with experimental values being much closer to the solid-liquid interface energy. The NN potential reduces a discrepancy by a third, relative to the EAM, but is still 20\%{} below the measured value at $\Tm$. 

\newcommand{\gsl}{\gamma_\text{sl}}
\newcommand{\musl}{\mu_\text{sl}}

\begin{figure}
    \includegraphics[width=1.\linewidth]{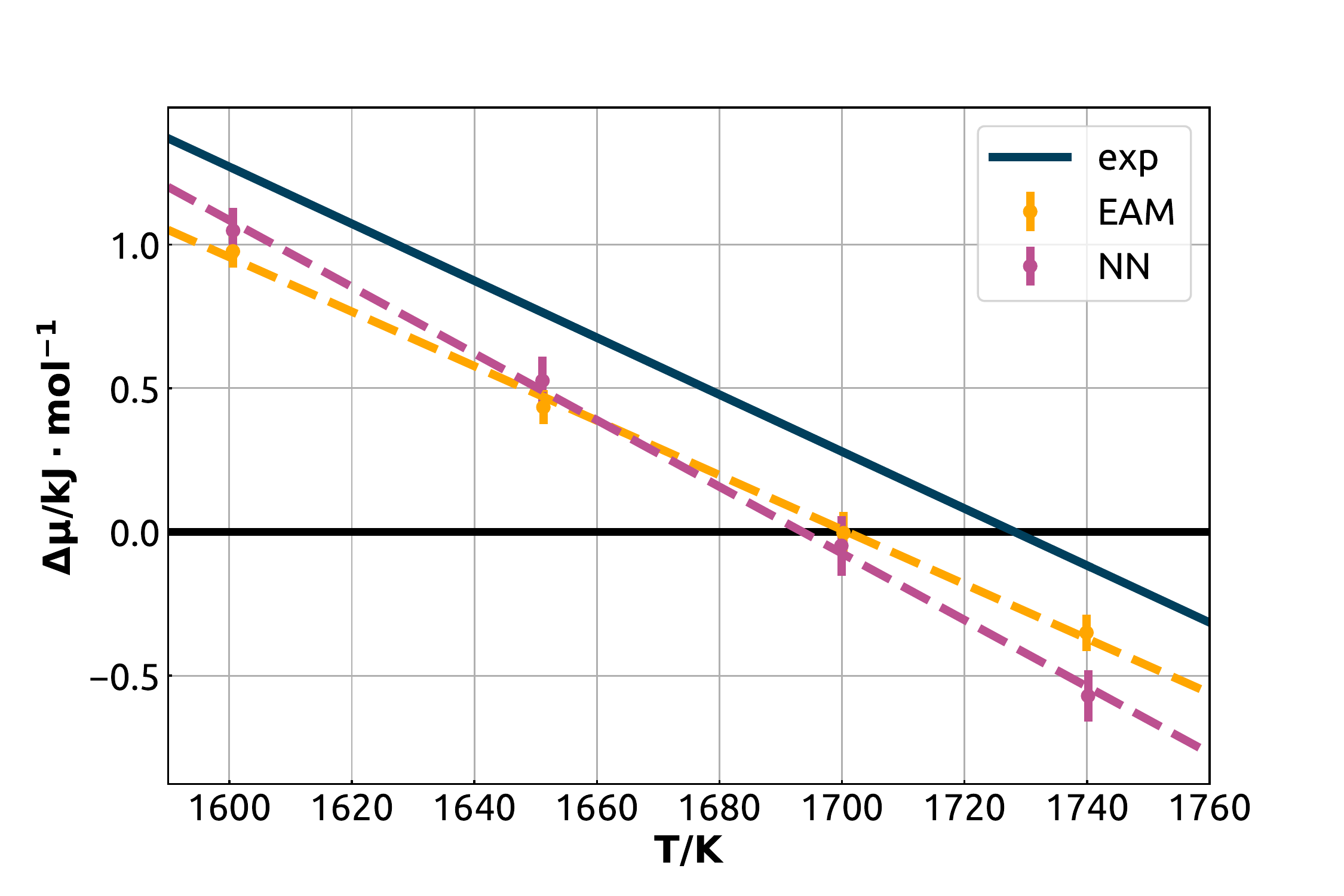}
    \caption{Chemical potential difference between solid and liquid phases of pure Ni as a function of temperature for EAM, NN and experiment \cite{chase1998nist}. The intersection of the yellow and purple lines with the black abscissa identifies the melting point for the corresponding potential.}
    \label{fig:tm}
\end{figure}

\subsubsection{Melting point and solid-liquid entropy}

Having separately characterized the properties of liquid and solid Ni at finite temperature, we can now turn to the determination of the relationship between the two phases and their interaction. We begin by characterizing the relative stability of the two phases and identifying their coexistence temperature. 

To this end, we use the interface pinning (IP) method\cite{pedersen_direct_2013} which works by applying a harmonic bias potential to a 2-phase system, which couples to an order-parameter $\Phi$ that discriminates between the two phases of interest. The Gibbs free energy difference between the phases is determined by the average force that the pinning potential exerts on the system.
As as order parameter to differentiate between solid and liquid we use the same collective variable discussed in Ref.~\cite{angi+10prb}, that uses cubic harmonics to identify environments that are \emph{fcc}-like and distinguish them from those that are liquid-like. 
If the mean value of the order parameter in the bulk solid and liquid at a given temperature is $\bar{\phi}_s$ and $\bar{\phi}_l$, and the sum over all atoms of the order parameter for a given configuration is $\Phi$, the number of solid atoms can be estimated as $N_s=(\Phi-N \phi_l)/(\bar{\phi}_s-\bar{\phi}_l)$ -- with the underlying assumption of choosing the dividing surface between the solid and the liquid phase that corresponds to zero excess for the chosen order parameter.~\cite{chen+15prb} 
With this definition, the Gibbs free energy associated with a two-phase configuration is given by $G(N_s) = \mu_s N_s + (N-N_s) \mu_l + 2\gsl A_{xy}$, where $\gsl$ is the solid-liquid interfacial free energy and $ A_{xy}$ the cross-section of the simulation box. When performing a simulation of the interface using a pinning potential of the form $(\Phi - \Phi_{ref})^2 \kappa/2$, the overall free energy reads
\begin{equation}
\tilde{G}(N_s) = \mu_s N_s + (N-N_s) \mu_l + 2\gsl A_{xy} + (\Phi - \Phi_{ref})^2 \kappa/2.\label{eq:gibbs-solidliquid}
\end{equation}

Hence, in conditions above or below the melting point the difference in chemical potential between the solid and the liquid phases leads to the interface fluctuating around an equilibrium position for which $\Phi\ne \Phi_{ref}$, and one can extract 
\begin{equation}
\Delta \musl = \mu_s - \mu_l = -\kappa (\Phi - \Phi_{ref}) (\bar{\phi}_s-\bar{\phi}_l)
\end{equation}
By performing multiple simulations at different temperatures one can identify the temperature dependence of $\Delta\musl$. The temperature at which $\Delta\musl=0$ identifies the melting point, and the slope is equal to the entropy of melting. As shown in Figure~\ref{fig:tm}, the computed melting points for EAM and NNP are 1700K and 1695K respectively -- only 2\% off the experimental value which is equal to 1728K. The slope of the two curves is also in good agreement with that of the experimental curve, corresponding to $\Delta S_\text{sl}=$ -9.48 (EAM) and -11.5 (NN) mJ/K, to be compared with the experimental value of -9.91 mJ/K. 

\begin{figure}
    \includegraphics[width=1.\columnwidth]{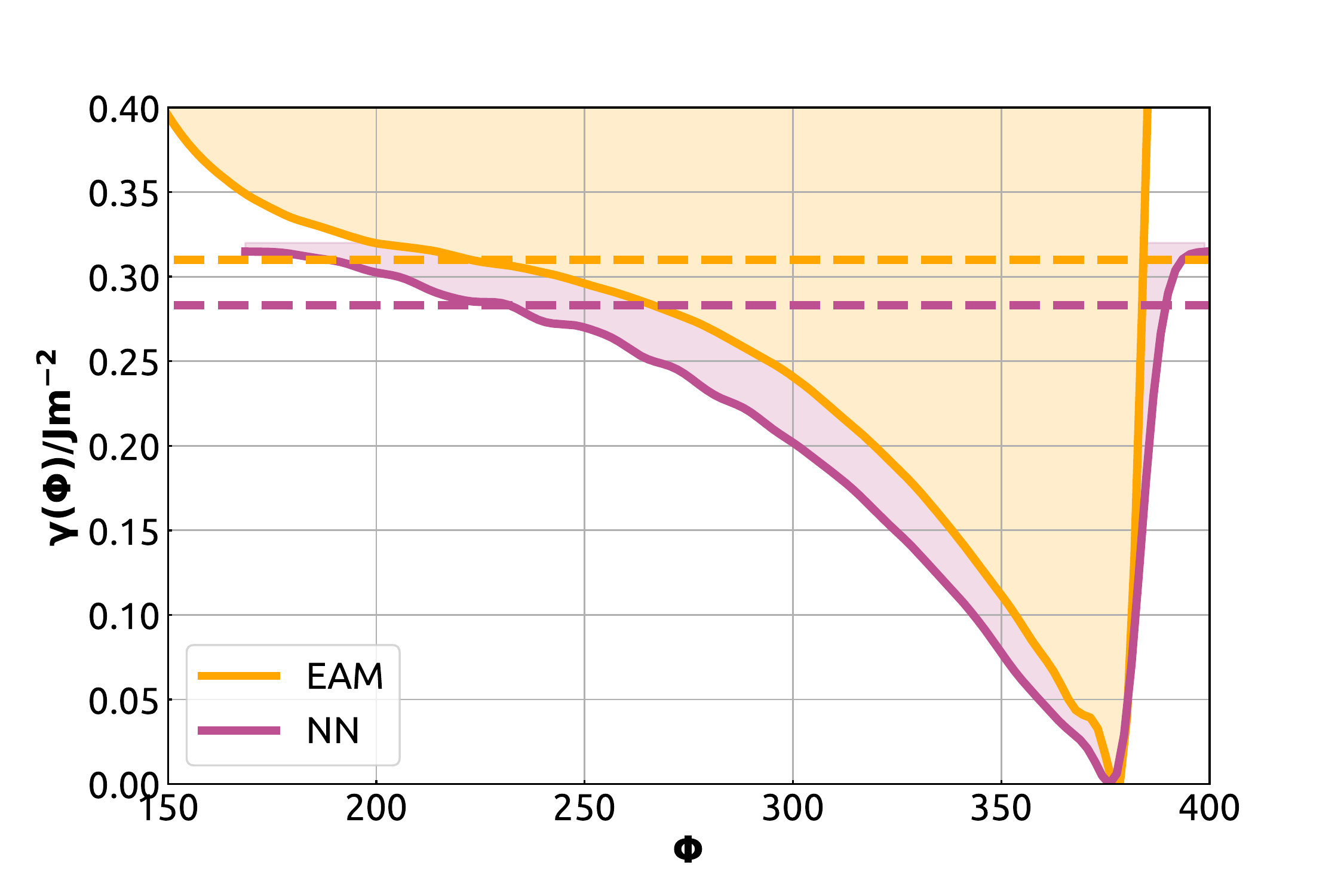}
    \caption{ The final free energy profiles $\gamma(CV)$ divided by the surface square and leveled with respect to the full solid state shown as a function of CV for EAM and NN. The profiles obtained during the metadynamic simulation of two-phase Ni system in a one dimensional CV space.}
    \label{fig:fes}
\end{figure}

\subsubsection{Solid liquid interface free energy}

The solid-liquid interface free energy plays a crucial role in determining the solidification behavior of materials, both in terms of controlling homogeneous nucleation, and in driving the formation of microstructure that in turns influences greatly the final materials properties.
Measuring $\gsl$ is however notoriously difficult, which triggered the development of several different methods to estimate it from atomistic modeling.\cite{hoyt_method_2001,broughton1986molecular, bai2006calculation}.
Here we use an approach that was first introduced in Ref.~\citenum{angi+10prb}, that relies on a bias potential to enable the reversible melting of a portion of an elongated simulation box (we use a box that is equivalent to $4\times 4\times 6$ \emph{fcc} unit cells, with the interface aligned along the (100) direction), and determine the constant $\gsl$ term in Eq.~\eqref{eq:gibbs-solidliquid} based on the free energy difference between a perfect solid and the configurations with two separate solid-liquid interfaces
\begin{equation}
    \gsl = \frac{G_{s(l)}-G_{s|l}}{2A_{xy}}.
\end{equation}
This expression is valid at $T=\Tm$, and for a planar interface -- whereas in out-of-equilibrium conditions~\cite{chen+15prb} or for a finite-size nucleus~\cite{chen-ceri18jcp} further subtleties arise including the dependency of the surface excess on the precise location of the solid-liquid dividing surface.

We build the bias that compensates for the interface free energy in an adaptive, history-dependent way, using the well-tempered metadynamics~\cite{laio-parr02pnas,bard+08prl} technique as implemented in PLUMED \cite{PLUMED,bonomi2019promoting}. Bias is built from repulsive Gaussians that are 0.007$eV$ high, have a width equal to 5 CV units (the same $\Phi$ order parameter used for the pinning poential) and that are added every 0.5ps. The well-tempered metadynamics bias factor ($\gamma=1+\frac{\Delta T}{T}$) is chosen to be 90. 
Given that, at the melting point, the depth of the well associated with the fully solid and the fully liquid states are equal, a restraint is also applied to restrict sampling and prevent complete melting. 
A sample PLUMED input is included in the SI.
As shown in Fig.~\ref{fig:fes} the free energy shows a minimum at large $\Phi$, corresponding to the fully-solid cell, and a plateau close to the restraining potential, corresponding to the presence of a solid/liquid interface. As observed in Ref.~\citenum{angi+10prb}, due to finite-size effects the free energy does not reach a clear-cut plateau when the solid-liquid interface is formed. This means that the value of $\gsl$ is affected by both a statistical and a systematic error - which we can estimate to be of the order of 0.03 based on benchmarks in Ref.~\citenum{angi+10prb}.
The free energy of this plateau makes it possible to estimate $\gsl=0.315$ and 0.283 Jm$^{-2}$ for EAM and the NN. The results are in a good agreement with previous calculations using investigated with the capillary fluctuation method: 0.234 Jm$^{-2}$  ~\cite{hoyt+01prl} and 0.325 Jm$^{-2}$ ~\cite{Rozas_2011} (calculated for the (100) surface); 0.287 Jm$^{-2}$ ~\cite{hoyt2004atoms} (averaged over different orientations).

\section{Conclusions}

In this work we demonstrate how the combination of machine-learning models and statistical sampling methods based on molecular dynamics makes it possible to compute the behavior of materials in realistic, finite-temperature conditions, including where necessary also the quantum mechanical nature of the nuclei.
We evaluate properties of the bulk solid and liquid phases, of defects, and of interfaces. 
The ML model makes it possible to achieve an accuracy on par with DFT, and even though for the specific case of Nickel excellent embedded atom potentials exist that reach comparable agreement with experiment as that obtained by the NNP, these results prove that it is possible to achieve predictive modeling of challenging thermodynamic properties without any experimental input, which shows great promise to facilitate the study of materials for which well-tested empirical potentials do not exist. 
What is more, we also show that a recently introduced ML model of the electron density can be used to incorporate a description of electronic excitations, which give a sizable contribution to the properties of Ni at and above the melting point. This serves as an example of the evolution of ML models from the construction of interatomic potentials to a more comprehensive replacement of quantum mechanical calculations, which brings one step closer the goal of fully predictive computational materials modeling and design. 

\begin{acknowledgments}
NL and MC acknowledge support by the CCMX project AM$^3$. CB acknowledges support by the Swiss National Science Foundation (Project No. 200021-182057).
\end{acknowledgments}

\end{document}